\documentclass[referee]{raa}
\usepackage{graphicx,times}
\usepackage{natbib}
\usepackage{amssymb,amsmath}
\usepackage{multirow}
\usepackage{threeparttable}
\usepackage{hyperref}
\hypersetup{pdftitle = The title of my PDF, pdfauthor = My name, pdfsubject= The subject, pdfkeywords = keyword1 keyword2 keyword3}
\hypersetup{colorlinks = true, linkcolor = green, anchorcolor = red, citecolor = blue, filecolor = red, urlcolor = red}

\begin{document}

\title{Study of the Modified Gaussian Model on olivine diagnostic spectral features and its applications in space weathering experiments}

 \volnopage{ {\bf 20XX} Vol.\ {\bf X} No. {\bf XX}, 000--000}
   \setcounter{page}{1}

   \author{Hui-Jie Han\inst{1}, Xiao-Ping Lu\inst{1}, Ya-Zhou Yang\inst{2}, Hao Zhang
   	\inst{3} \inst{4}, Admire Muchimamui Mutelo\inst{5}
   }

   \institute{ State Key Laboratory of Lunar and Planetary Sciences, Macau University of Science and Technology, Taipa 999078, Macau, China; {\it xplu@must.edu.mo}\\
	\and
State Key Laboratory of Space Weather, National Space Science Center, CAS, Beijing 100190, China\\
	\and
Planetary Science Institute, School of Earth Sciences, China University of Geosciences, Wuhan 430074, China\\
	\and
CAS Center for Excellence in Comparative Planetology, Hefei 230026, China\\
	\and
School of Earth Science and Resources, China University of Geosciences, Beijing 100083, China\\
\vs \no
}

\abstract{The absorption features of olivine in visible and near-infrared (VNIR) reflectance spectra are the key spectral parameters in its mineralogical studies. Generally, these spectral parameters can be obtained by exploiting the Modified Gaussian Model (MGM) with a proper continuum removal. However, different continua may change the deconvolution results of these parameters. This paper investigates the diagnostic spectral features of olivine with diverse chemical compositions. Four different continuum removal methods with MGM for getting the deconvolution results are presented and the regression equations for predicting the Mg-number (Fo\#) are introduced. The results show that different continua superimposed on the mineral absorption features will make the absorption center shift, as well as the obvious alterations in shape, width, and strength of the absorption band. Additionally, it is also found that the logarithm of a second-order polynomial continuum can match the overall shape of the spectrum in logarithmic space, and the improved regression equations applied to estimate the chemical composition of olivine-dominated spectra also have a better performance. As an application example, the improved approach is applied to pulse laser irradiated olivine grains to simulate and study the space weathering effects on olivine diagnostic spectral features. The experiments confirm that space weathering can make the absorption band center shift toward longer wavelength. Therefore, the Fo\# estimated from remote sensing spectra may be less than its actual chemical composition. These results may provide valuable information for revealing the difference between the spectra of olivine grains and olivine-dominated asteroids.	
}

   \authorrunning{H.-J. Han et al. }            
   \titlerunning{Study of the MGM and its applications}  
   \maketitle

%
\section{Introduction}           
\label{sect:introd}

Hyper spectral remote sensing data are widely used in identification of minerals and rocks (\citealt{adams1974visible, Singer1981Near, cloutis2000diaspores, burbine2002spectra, basilevsky2012geologic, Roush2015Laboratory}), especially, on studies of the surface materials of the Moon, Mars, and asteroids (\citealt{Pieters1988Exploration, bishop1998spectroscopic, ohtake2009global, Wu2010Global, lindsay2015composition}). The visible and near-infrared (VNIR) reflectance spectra are sensitive to the mafic mineral compositions (\citealt{lucey2004mineral, Staid2011The}). As a major mafic mineral, olivine is an important material in understanding the geologic evolutionary process of terrestrial bodies. It is the foremost mineral to crystallize from the magma and form the major component of planetary mantles. Olivine bears rich information for the magma ocean. Olivine materials in the deep crust and mantle can be exposed to the crust or surface by the magmatic explosion and giant impacts (\citealt{Tonks1992Magma, wilson2017eruption}), and detected by remote sensing instrument. Furthermore, it was found that high Fo\# (100$\times$Mg/(Mg+Fe)) olivine indicates a primordial magma of mantle origin, while low Fo\# olivine indicates a differentiated and evolved magma (\citealt{Taylor1978Geochemical, Shearer1999Magmatic}). The composition of olivine is an indicator to study the mantle materials and planetary evolutionary history.

\subsection{Olivine spectroscopy}
\label{subsec:Olivine spectroscopy}

Visible and near-infrared spectrometer (VNIS) is generally an essential instrument for orbital remote sensing. Generally, the VNIR spectra of absorbing minerals contain absorption bands that are characteristic of their compositions and crystal structures (e.g., \citealt{mustard2005olivine, pieters2009moon}). The crystalline structure of olivine contains two different six-coordinated sites: the centrosymmetric Ml site and non-centrosymmetric M2 site (\citealt{burns1970crystal}). In the series from fayalite to forsterite, Mg$^{2+}$ and Fe$^{2+}$ ions are randomly distributed within the two sites, and the Fe$^{2+}$/ Mg$^{2+}$ ratio can give information on the differentiation process of the olivine (\citealt{burns1974polarized, dyar2009spectroscopic}). The presences of Fe$^{2+}$ in the central absorption site M2 and two bilateral absorption site M1-1 \& M1-2 of the olivine crystal field make three spin-allowed transition and form three corresponding characteristic absorption features in the VNIR spectrum near 0.87, 1.05, and 1.25 $\mu$m (\citealt{hunt1977spectral, hunt1979spectra}). The combination of these absorption bands result in a broad integrated diagnostic feature near 1.0 $\mu$m (Fig. 1), and its useful spectral behavior finally makes olivine chosen for this study.

Identification of mafic mineral olivine is done by composition analysis of laboratory mineral samples and via satellite remote sensing detection (e.g., \citealt{mustard2005olivine}). The scientific study of olivine content, crystal structure, and spectral features, will help us to uncover the formation and evolution of the planets and moons (e.g., \citealt{lucey1998feo, Yamamoto2010Possible, miyazaki2013olivine, ody2013global, li2019chang}). Remote sensing can acquire a large amount of spectral information. However, the spectral data are not directly exploited, since the spectral analysis requires much prior knowledge with many attempts in the correction for absorption bands of minerals to fit different models (\citealt{antonenko2003analysis}).

\subsection{The modified Gaussian model (MGM)}
\label{subsec:The modified Gaussian model (MGM)}

The modified Gaussian model (MGM), originally developed and described by \cite{Sunshine1990Deconvolution}, is based on physical processes in electronic transition absorptions and statistical method of the probability distribution. It is an automated mathematic method for deconvolving the superposition absorption features in reflectance spectra. MGM is a widely used model for accurately decomposing the mineral absorption features into their physical isolated absorption bands, and can estimate the single mineral modal abundances within 5\% $\sim$ 10\% for mafic minerals and rocks (\citealt{Sunshine1993Estimating}). It has been successfully used to interpret and quantify the compositional and abundance information on minerals or mineral mixtures, which make up the observed surface (e.g., \citealt{Sunshine1998Determining, mustard2005olivine, jin2013new}). Many studies have applied and adapted the method based on experience and detailed explanation of the parameters of the resulting absorption band processed by this model (e.g., \citealt{Pinet2009Mafic, Tsuboi2010A, Sugita2011A}).

It is worth noting that using the MGM to analyze the spectra in wavenumber domain (spatial frequency or energy) is numerically equivalent to Gaussian analysis in wavelength (\citealt{gallie2008equivalence}). In addition, it is necessary to pay attention to the constraints between the relative band absorption strengths. Each Gaussian curve must conform to the electronic transition absorption laws, while getting the optimal numerical solution. The MGM code applied in this paper is the version 2.0 MATLAB code, which calculates the best-fit bands using the stochastic non-linear inverse algorithm developed by \cite{Tarantola1982Generalized}. The code can be downloaded from the Brown University $\footnote{\url{http://www.planetary.brown.edu/mgm/}}$.

MGM is very useful for the planetary remote sensing data interpretations. Continuum removal as an important issue in spectral interpretation is still not been analyzed in detail with the MGM. Indeed, the continuum is superimposed on the mineral absorption features, which can cause the absorption center to shift, as well as the obvious alterations in shape, width, and strength of the absorption band (\citealt{Rodger2012A, Zhang2016Study}). However, it seems to be using markedly different continuum line styles in the MGM deconvolution in \cite{Sunshine1998Determining} and others (e.g., \citealt{noble2006using, clenet2011new, li2019chang}). Different continua will make a change of the band center. In addition, the MGM deconvolution relies on relative band strength values and band center positions. Therefore, continua of different slopes and types will lead to different results for MGM deconvolution. So, it is necessary to study the effects of different continuum removal methods on MGM.

\subsection{MGM and space weathering}
\label{subsec:MGM using in space weathering spectra}
Space weathering is the surface process that can cause physical and chemical changes on the surface of airless bodies (\citealt{hapke2001space}). This process, including micrometeorite impacts and solar wind implantations, can darken, redden and reduce the contrast of the VNIR spectra of most silicate minerals, and make the spectral difficult to interpret (\citealt{Pieters1993Optical, chapman2004space}). Researches have pointed out that “nanophase iron particles”, abbreviated as npFe$^{0}$, disassociation from iron-bearing minerals, produced by irradiation and vapor deposition effects, can cause the changes of optical properties and chemical composition on the material surface (\citealt{pieters2000space, hapke2001space, brunetto2006space, fu2012effects, Yang2017Optical}).

At present, pulsed laser irradiation and ion bombardment are the common experimental simulation methods for space weathering simulation (e.g., \citealt{kohout2014space}). It has been used to simulate space weathering on olivine grains for interpreting the material composition of lunar soil, A-type asteroids, and S-type asteroids(e.g., \citealt{chapman1996s, pieters2000space, chapman2004space}). MGM is widely used to reveal the surface mineralogy of asteroids (e.g., \citealt{Sunshine2004High, Sunshine2007Olivine, deLeon2004mineralogical, binzel2009spectral,Wang2016Analysis}). However, space weathering will complicate the interpretations of telescopic spectra of asteroids (\citealt{clark2002asteroid, chapman2004space}). By far, only a handful of studies optimized the effects of space weathering on MGM deconvolution results. For example, \cite{fu2012effects} studied the effects of space weathering on mineral diagnostic spectral features based on the results of He$^{+}$ irradiation experiments and pointed out that the diagnostic band centers shift to longer wavelength after radiation. There is a need for more experiments based on the application of MGM approach in space weathering.

In this work, firstly we collected spectra data of olivine with different particle sizes and compositions, solid solution ranging from fayalite (Fe$_{2}$SiO$_{4}$) to forsterite (Mg$_{2}$SiO$_{4}$), and then examined different continuum removal methods used for MGM analysis. Secondly, we studied the effects of space weathering on MGM analysis using the spectra of laser-irradiated olivine and olivine-dominated A-type asteroids Asporina (246) and Eleonora (354).

\section{Data and methods}
\label{sect:Data and methods}

\subsection{VNIS spectra and composition of the olivine solid solution}
\label{subsec:VNIS spectra and composition of the olivine solid solution}
The spectral laboratory data described here are provided by the U.S. Geological Survey (USGS) $\footnote{\url{https://speclab.cr.usgs.gov/spectral-lib.html}}$ and RELAB Spectral Database is supported by NASA $\footnote{\url{http://www.planetary.brown.edu/relab/}}$. The olivine spectra were measured from 446 to ~2700 nm (some measuring endpoint value is 2652 nm or 2682 nm) in the USGS spectral laboratory and 300 to 2600 nm from RELAB database. The USGS M3t data convolved the mineral spectral data to the target mode of the Moon Mineralogy Mapper (M$^{3}$) spectrometer with spectral characteristics determined by \cite{green2011moon} (\citealt{kokaly2017usgs}).

The spectra of olivine with several different size distributions are used in this study. The olivine sample ID and information are listed in Table 1. Seven samples, ID likes KIxxxx, have been also used in \cite{Sunshine1998Determining}. One should note that the Fo\# was recalculated according to the chemical composition of the USGS olivine samples. It established that the Fo\# are a little different from the sample title. For accuracy, the recalculated Fo\# was used in this study. Olivine spectra in RELAB Spectral Database are synthetic samples, as described by \cite{dyar2009spectroscopic}, and the Fo\# is an approximation of a range.

\begin{figure}
	\centering
	\includegraphics[width=12.0cm, angle=0]{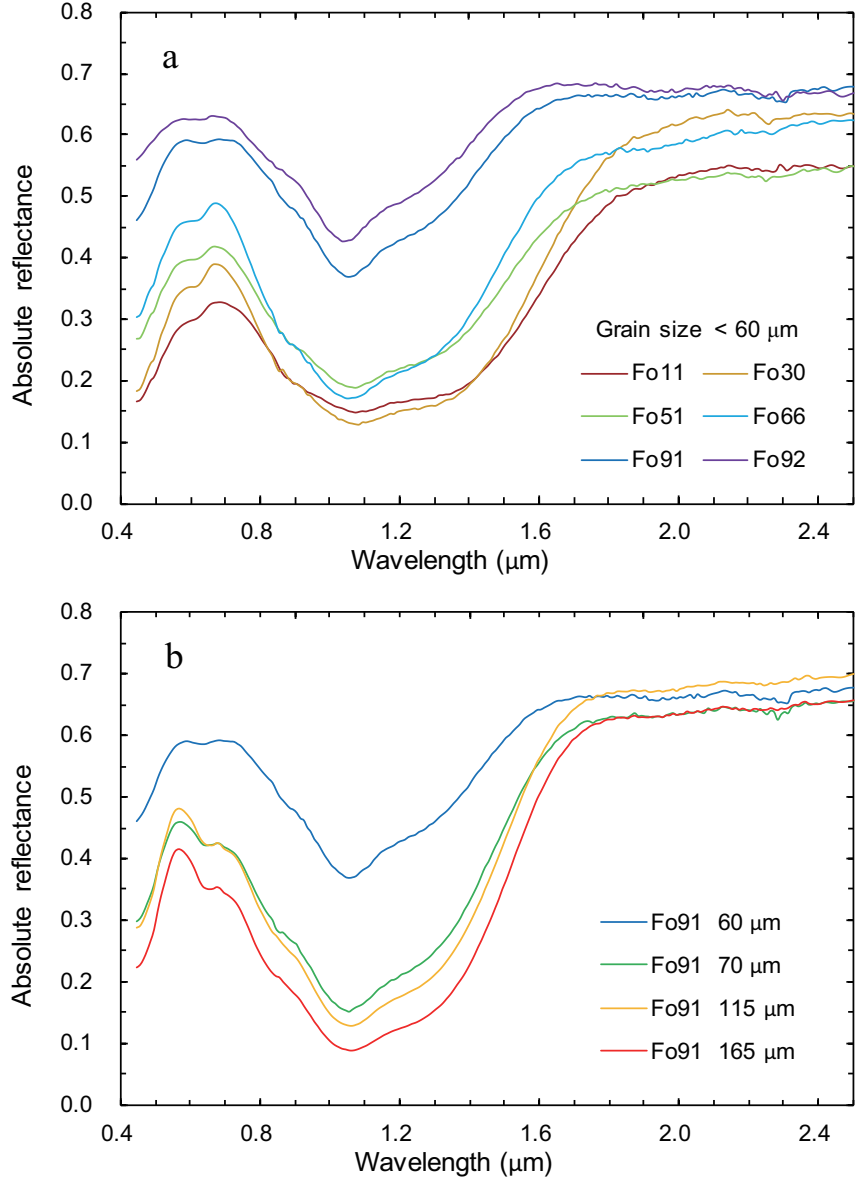}
	\caption{Reflectance spectra of olivine. Diagram a: Olivine of varied Fo\# with same grain size, all spectra are for  $<$ 60  $\mu$m size powders, showing increasing reflectance and decreasing band depth with increasing Fo\#; Diagram b: Olivine spectra of varied Fo\# with varied grain size (60 to 165  $ \mu$m), showing decreasing reflectance with increasing grain size.}
	\label{Fig1}
\end{figure}

These USGS olivine can be divided into two groups as shown in Fig. 1: a) they have the same grain size ($<$ 60 $ \mu$m) with increasing Fo\#; b) they have the same Fo\# with increasing grain size (60 $ \mu$m to 165 $ \mu$m). Usually, laboratory spectra with larger grain size ($>$ 45 $ \mu$m) are more similar to the natural mineral spectra than the small grain size (\citealt{clenet2011new}). Making it with the same grain size, powder size $<$ 60 $ \mu$m, the spectral reflectance increases and band depth decreases with increasing Fo\# (decreasing Fe$^{2+}$). Obviously, as the iron abundance increases, the band center position of the 1.0 $\mu$m absorption band shifts toward longer wavelength. In addition, the particle size also affects the spectra reflectance and band shape (\citealt{Salisbury1985The, SALISBURY1992The}). The spectral absorption features of forsterite shifts toward longer wavelengths when the grain size increases. When the band saturation in very large grain sizes occurs, the band center determination will become complicated.

\begin{table}
	\bc
	\begin{minipage}[]{150mm}
		\caption[]{Olivine spectra selected for studying continuum removal methods in this study
		\label{tab1}}\end{minipage}
	\setlength{\tabcolsep}{2.0mm}
	\small
	\begin{tabular}{ccccccccccc}
		\hline\noalign{\smallskip}
	\multicolumn{3}{c}{USGS Spectral Library\tnote{1}}	&& \multicolumn{7}{c}{RELAB Spectral Database\tnote{2}}\\
		\cline{1-3} \cline{5-11}
	Sample ID&  Size($\mu$m) &Fo\#& 	&Sample ID  &Size($\mu$m)&Fo\#& 	&Sample ID  &Size($\mu$m)&Fo\#\\
		\cline{1-3} \cline{5-7} \cline{9-11}
	KI3005& 	$<$ 60	&11&		&DD-MDD-087	&$<$ 45	&80&		&DD-MDD-115	&$<$ 45	&89.5\\
	KI3377&		$<$ 60	&19&		&DD-MDD-088	&$<$ 45	&75&		&DD-MDD-116	&$<$ 45	&70\\
	KI3291&		$<$ 60	&30&		&DD-MDD-089	&$<$ 45	&70&		&DD-MDD-046	&$<$ 45	&0\\
	KI4134&		$<$ 60	&41&		&DD-MDD-090	&$<$ 45	&65&		&DD-MDD-045	&$<$ 45	&10\\
	KI3188&		$<$ 60	&51&		&DD-MDD-091	&$<$ 45	&60&		&DD-MDD-044	&$<$ 45	&20\\
	KI3189&		$<$ 60	&60&		&DD-MDD-092	&$<$ 45	&55&		&DD-MDD-043	&$<$ 45	&30\\
	KI3054&		$<$ 60	&66&		&DD-MDD-093	&$<$ 45	&50&		&DD-MDD-042	&$<$ 45	&40\\
	GDS70.d&	$<$ 60	&91&		&DD-MDD-094	&$<$ 45	&40&		&DD-MDD-041	&$<$ 45	&50\\
	GDS71.b&	$<$ 60	&92&		&DD-MDD-095	&$<$ 45	&30&		&DD-MDD-040	&$<$ 45	&60\\
	GDS70.e&	$<$ 30	&91&		&DD-MDD-096	&$<$ 45	&20&		&DD-MDD-039	&$<$ 45	&70\\
	GDS70.c&	$<$ 70	&91&		&DD-MDD-097	&$<$ 45	&10&		&DD-MDD-038	&$<$ 45	&80\\
	GDS70.b&	$<$ 115	&91&		&DD-MDD-098	&$<$ 45	&0&			&AG-TJM-008	&$<$ 45	&90\\
	GDS70.a&	$<$ 165	&91	\\	
		\noalign{\smallskip}\hline
	\end{tabular}

	\ec
	\tablecomments{0.95\textwidth}{The origins of the samples involved in this paper are different. Olivine in USGS Spectral Library are terrestrial olivine in nature, and their Fo\# are accurately calculate to the unit digit with the chemical composition analysis (e.g., KI3054, Fo66 is an approximation of the exact number 65.53). Olivine in RELAB Spectral Database analysis (e.g., KI3054, Fo66 is an approximation of the exact number 65.53). Olivine in RELAB Spectral Database are synthetic samples, as described by \cite{dyar2009spectroscopic}, and their Fo\# are approximation of a range (e.g., DD-MDD-091, Fo60 is an approximation number ranging from 53 to 64).}
	
\end{table}

\subsection{Using MGM to deconvolve olivine spectra}
\label{subsec:Using}
In MGM, natural logarithm of each spectral reflectance is modelled as:
\begin{equation}
\label{Eqn1}
\ln\left [ R\left ( \lambda  \right ) \right ]=F\left ( cont \right )+ \sum_{i=1}^{N}S_{i}\cdot \exp\left [ -\frac{\left ( \lambda ^{-1}-\mu _{i}^{-1} \right )^{2}}{2\sigma _{i}^{2}} \right ]
\end{equation}

where $\lambda$ is the wavelength, \emph{R} ($\lambda$) is the reflectance at wavelength $\lambda$, \emph{S} is the band strength, $\mu$ is the band center, $\sigma$ is the band width, and \emph{N} is the number of bands. \emph{F}(\emph{cont}) is the function of the continuum in wavenumber space.

The MGM is not very dependent on pre-hypothesis of spectral components, but band absorption analysis needs to be done with suitable caution to ensure that the absorption have physical significance. In the olivine spectra, there are two inflection regions near 700 and 1800 nm, and a wide absorption near 1 $ \mu $m region. The Gaussian curves in Fig. 2 indicate that there are three absorptions (870 nm, 1050 nm, and 1250 nm, respectively, which are assigned to transitions of Fe$^{2+}$ in the M1-1, M2, and M1-2 site of olivine) in this region (\citealt{Sunshine1998Determining}). The initial parameters based on prior knowledge of the mineral spectral absorption can be set in advance when running the MGM code. Therefore, the number of Gaussians curves to be used in the model directly depends on the number of absorption bands present in the spectrum. In order to elaborate on this model, there are four basic characteristics significant for interpreting and exploiting the MGM, as shown in Fig. 2. They are: 1) band center (the central location of each absorption feature); 2) band width (typically described as the full width at half maximum (FWHM) in the independent reflectance spectrum curve); 3) band strength (the band intensity in natural logarithmic reflectance space); and 4) continuum (a mathematical function used to prominence to show a particular spectral absorption feature for analysis) (\citealt{Sunshine1993Estimating, clark1984reflectance}).

\begin{figure}
	\centering
	\includegraphics[width=12.0cm, angle=0]{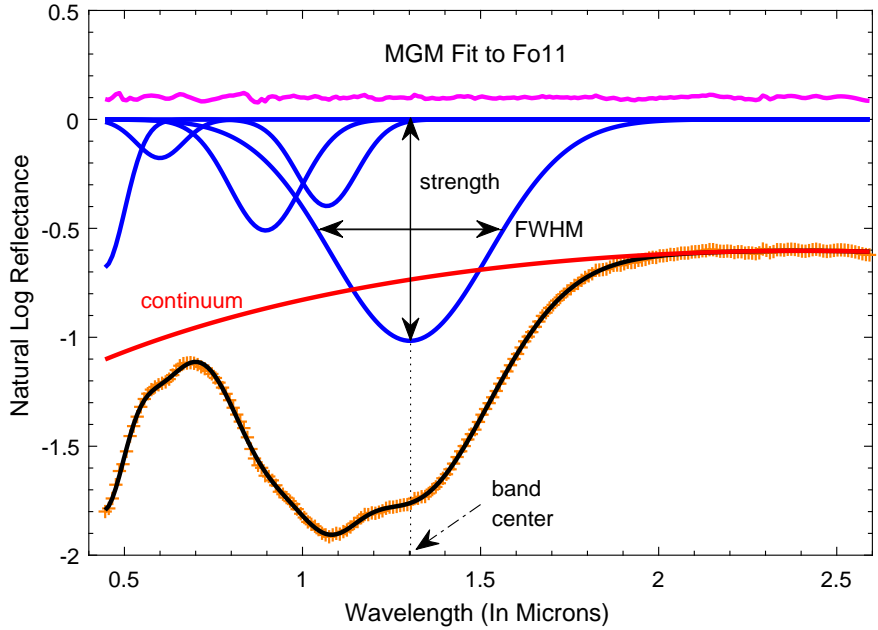}
	\caption{Four significant characteristic parameters for MGM. In this figure, the spectral data (Sample ID: KI3005, Fo11) is from the USGS spectral library. The orange “+” points are the logarithm reflectance data of KI3005. The blue line is the Gaussian curves fitting for the reflectance data. The red line represents a type of continuum used in the MGM. The black line is the fitting result represents the Gaussian curves plus the continuum. The pink line represents the residuals (original data minus modeled Gaussian fitting result), for the sake of clarity, the residuals are shifted by +0.1 on the ordinate.}
	\label{Fig2}
\end{figure}

\subsection{Different continuum removal methods}
\label{subsec:Different}
The slope of the continuum is variable, which consists of a number of factors such as grain size, viewing geometry, space weathering, temperature, et al. The variable continuum is the barrier for the quantitative analysis of spectral diagnostic absorption features (\citealt{isaacson2011remote}). To quantify the absorption band features, it is necessary to remove the continuum from the spectral curve, however, the physical significance of the continuum is not thoroughly understood yet (\citealt{clark2003imaging, clenet2011new}). Accurately modeling a suitable “continuum” with superimposed absorptions is not easy but crucial for analyzing the spectroscopic data with the MGM.

In the MGM program, users can set clear continuum polynomial parameters. In order to compare the relative absorption band intensity and study the effect of the continuum removal method (short for C) on the MGM deconvolution, 4 different types of continuum removal methods  in wavelength space are used in this article to get the olivine diagnostic spectral features:
1. Flat line continuum removal method (short for Method F)\\
Use the logarithm of the max reflectance values and one additional model parameter of offset as a flat continuum to remove in wavelength space.

\begin{equation}
\label{Eqn2}
C\left ( \lambda  \right )=\ln\left ( MAX_{\left ( R \right )} \right )+c_{0}
\end{equation}

\begin{center}
\emph{R}= reflectance, $\lambda$= wavelength, $c_{0}$= constant
\end{center}

2. Oblique line continuum removal method (short for Method O)\\
Use the logarithm of first-order trend line as a linear equation tangent continuum to remove in wavelength space.

\begin{equation}
\label{Eqn3}
C\left ( \lambda  \right )=\ln\left ( c_{0} +c_{1} \lambda\right )
\end{equation}

\begin{center}
 $\lambda$= wavelength,  $c_{0}, c_{1}$= constants
\end{center}

3. Polynomial curve continuum removal method (short for Method P)\\
Use the logarithm of second-order polynomial curve as the continuum to remove in wavelength space.

\begin{equation}
\label{Eqn4}
C\left ( \lambda  \right )=\ln\left ( c_{0} +c_{1} \lambda+c_{2} \lambda^{2}\right )
\end{equation}

\begin{center}
	$\lambda$= wavelength,  $c_{0}, c_{1}, c_{2}$= constants
\end{center}

4. Energy- wavelength polynomial continuum removal method (\citealt{hiroi2000improved}) (short for Method E)
Use the energy-wavelength polynomial as the continuum to remove in wavelength space.

\begin{equation}
\label{Eqn5}
C\left ( \lambda  \right )= \frac{c_{0}}{\lambda} +c_{1} +c_{2} \lambda
\end{equation}

\begin{center}
	$\lambda$= wavelength,  $c_{0}, c_{1}, c_{2}$= constants
\end{center}

\begin{figure}
	\centering
		\begin{minipage}{7cm}
			\centering
			\includegraphics[width=7.2cm]{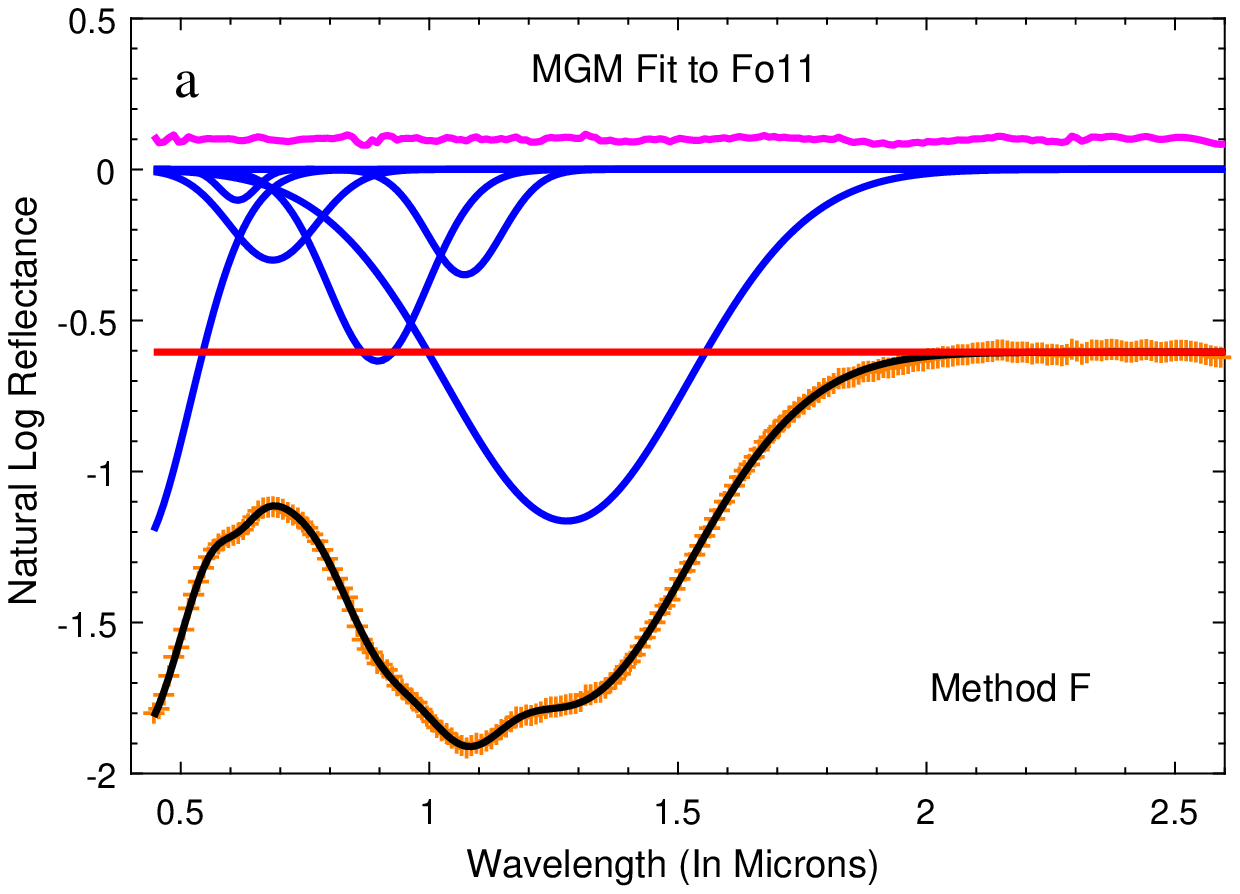}
		\end{minipage}
		\begin{minipage}{7cm}
			\centering
			\includegraphics[width=7.2cm]{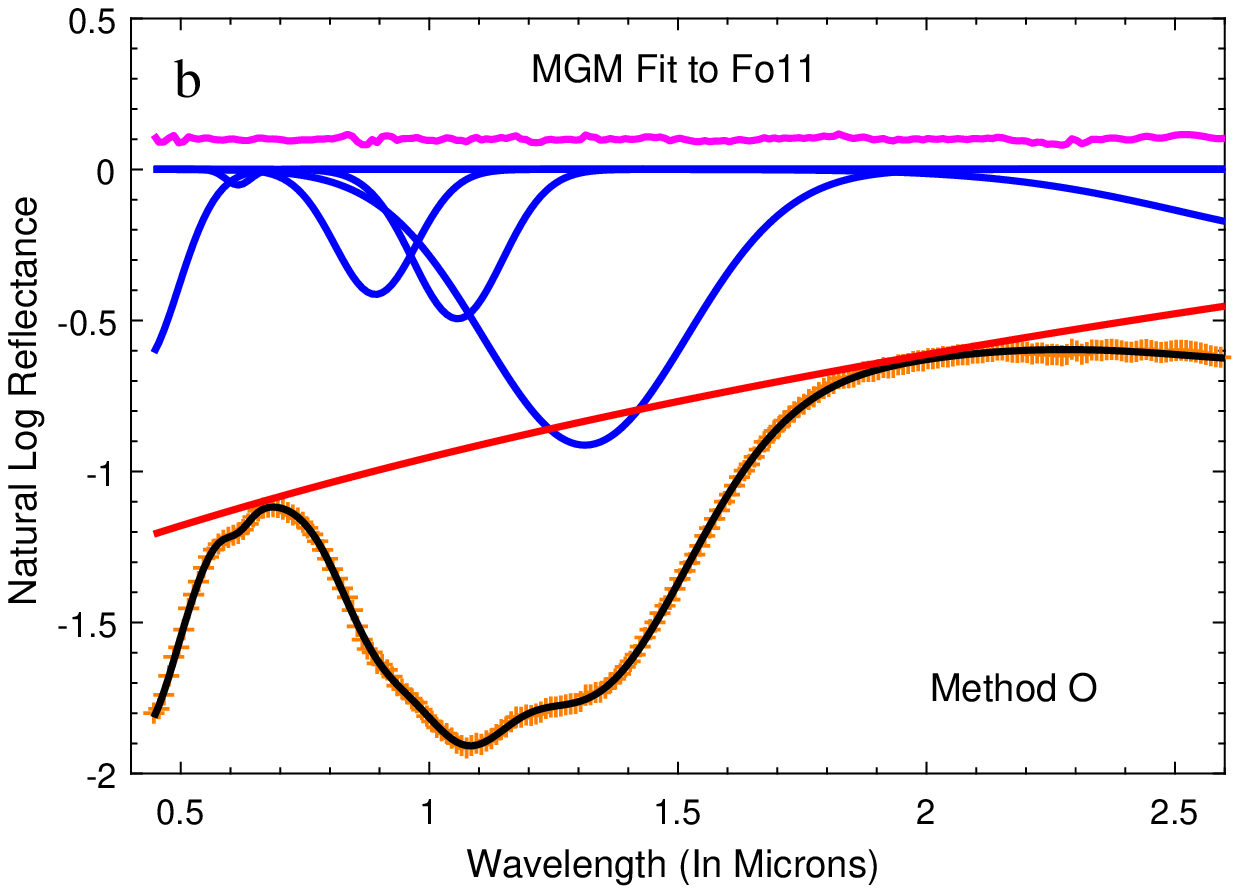}
		\end{minipage}
		\begin{minipage}{7cm}
			\centering
			\includegraphics[width=7.2cm]{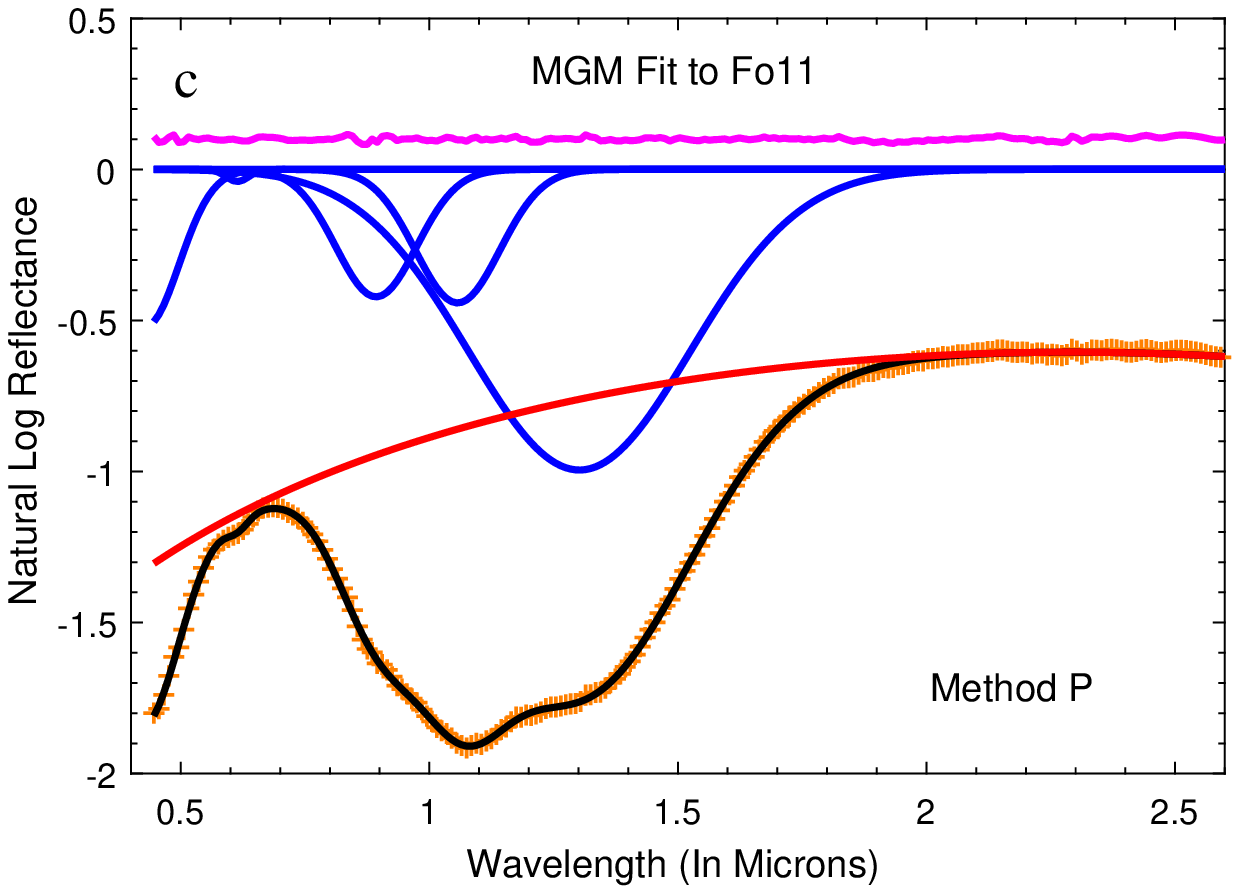}
		\end{minipage}
		\begin{minipage}{7cm}
			\centering
			\includegraphics[width=7.2cm]{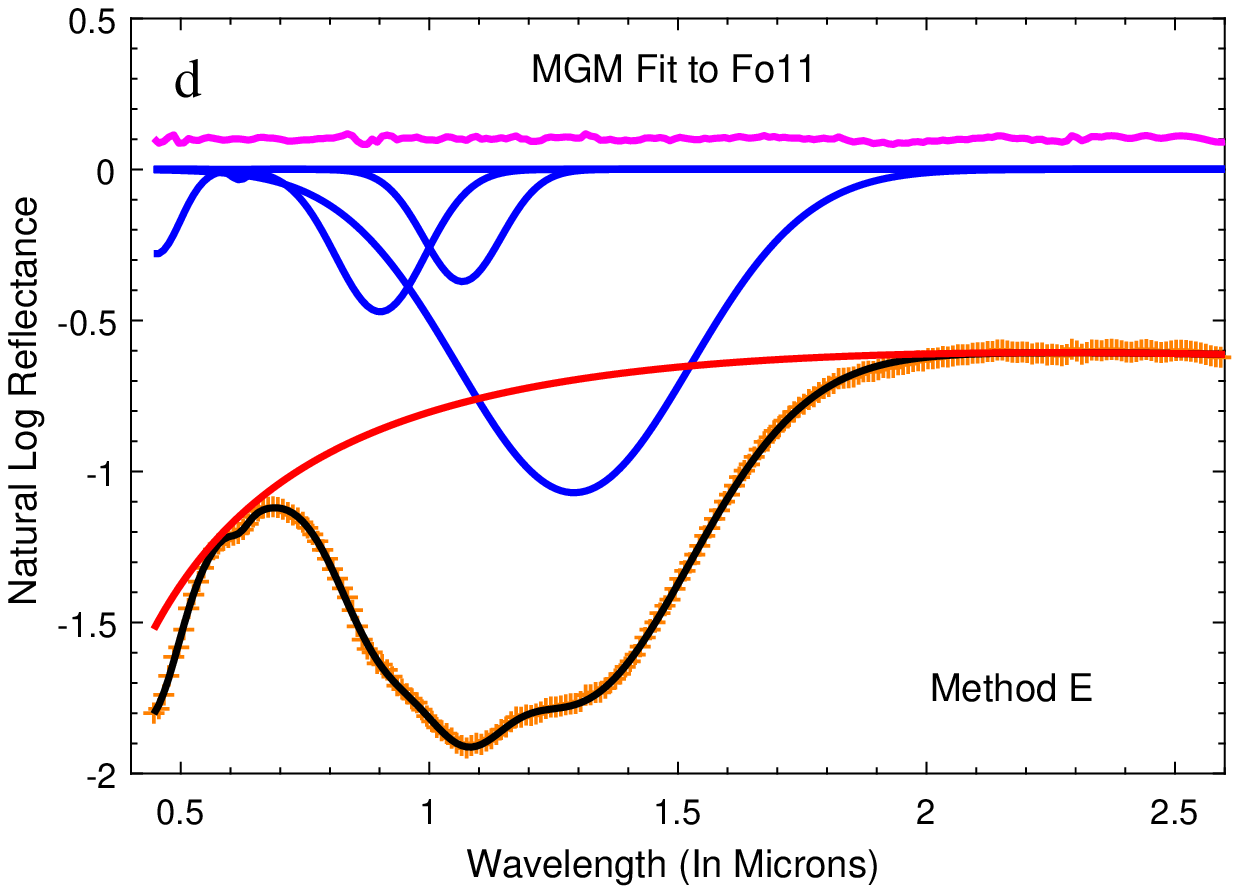}
		\end{minipage}
	\caption{Olivine (Fo11) spectral deconvolution results from four continuum removal methods of MGM. Diagram a, b, c, and d represent the four methods of F, O, P, and E, respectively. a: flat line continuum without slopes was used to match the flatness spectra in the long wavelength regions for simplicity. b: oblique line continuum tangent to the two inflection regions on both sides of the absorption region. c: quadratic polynomial curve continuum describes the general spectral trend. d: energy-wavelength polynomial continuum describes the general spectral trend.}
	\label{Fig3}
\end{figure}

Flat line, oblique line, and polynomial curve continuum removal methods are commonly used in the MGM deconvolution. \cite{hiroi2000improved} developed Method E for deconvolving the reflectance spectra of lunar soils. Fig. 3 shows an example of MGM fit to fayalite (Fo11) using the above mentioned four continuum removal methods.

\subsection{MGM parameters and initialization}
\label{subsec:MGM}

Following the previous explanation in Section 2.2 and 2.3, each Gaussian curve must be initialized with three parameters (band center, band width, and band strength) and each parameter corresponds to a 2$\sigma$ uncertainty. Likewise, the continuum must be initialized as a certain type with the corresponding 2$\sigma$ uncertainty for each parameter. In addition to the 1 $\mu$m region, sometimes, another two or more Gaussians at the short wavelength region, and one or more Gaussians at the long wavelength region are also needed to obtain physically realistic modeling to initialize the MGM. The former is used to model the strong large absorption of charge transfer near ultraviolet region and weak absorption near 0.6 $\mu$m (\citealt{clenet2011new}), and the latter are used to model the absorption of atmospheric water in the long-wavelength region (\citealt{dyar2009spectroscopic}).

The  transmission spectra of olivine have been studied in detail by \cite{burns1970crystal}. In his work, spectra from all three ($\alpha, \beta, \gamma$) crystallographic optical orientations were examined as a function of compositions. He deconvolved each of the $\alpha, \beta,  \gamma$ polarized spectra into three absorption bands. In the $\alpha$ and $\beta$ polarized spectra, M1-1 band and M1-2 band contribute most strongly to the spectra, and M2 band has little or no contribution to the spectra. Whereas, M2 bands are most intense in the $\gamma$ orientation, and form a narrow and sharp curve. However, reflectance spectra of pulverized samples are different from the polarized transmission spectra of single crystals in that the former ones include scattering contributions from randomly orientated micro-crystal. The shape of reflectance spectra is more similar to $\alpha$ and $\beta$ polarized spectra, but not the sharp curve of the $\gamma$ polarized spectra. To deconvolve olivine spectral absorptions more accurately, the research presented here initialize the intensity of M2 to about half of M1-2.

The relative strength shifts with the variations of olivine Fo\#, although this relationship is less useful for compositional analysis, it should be also concerned as mentioned in \cite{Sunshine1998Determining}. A series of tests of the MGM modeling have driven us to set appropriate Gaussian parameters and determine the uncertainty of 50 nm for the band center, 50 nm for the band width, and within 30\% for the band strength. For the spectra of poor quality, it can also reduce the 2$\sigma$ uncertainty to model the data accurately. Continuum of Method O, P and E were modeled as close as possible to the two inflection regions near 700 nm and 1800 nm. The parameters of the continuum are variable with different selected methods, and we set an uncertainty within 30\% for each continuum parameter.

\subsection{Space weathering simulations of olivine grains}
\label{subsec:Space}

In order to test how MGM may work for samples experienced simulated space weathering, we re-analyzed the reflectance spectra of olivine grains (less than 45 $\mu$m) irradiated by pulse laser with different density levels (\citealt{Yang2017Optical}). The detailed experimental procedures can be found in \cite{Yang2017Optical} and the following is a brief summary:

In order to study the effects of space weathering on olivine diagnostic spectral features, some natural pure olivine granules were collected and ground into powder (less than 45 $\mu$m) as the sample. The wet chemistry method was used to analyze the major contents of olivine elements and its Mg-number is Fo89.8. Olivine powders were baked with a dry oven under 120 $^{\circ}$C for more than 5 hours to remove any moisture in the samples, and then the dry samples were placed in aluminum holders in a vacuum chamber under a pressure of 10$^{-3}$ Pa for pulsed laser irradiation. By varying the pulse energy and irradiation time, it can simulate different weathering degrees.

We obtained totally 6 olivine samples with varied weathering degrees, including the original unirradiated one. A Continuum nanosecond pulsed laser was used to simulate the micrometeorite bombardment process. The pulse energy was set as 25 mJ/pulse, and different irradiation times were used for these samples to simulate varied degrees of micrometeorite bombardments. Before and after irradiations, VNIR reflectance spectra of these olivine samples were measured with a Bruker Vertex 70 Fourier transform infrared spectrometer. The transmission electron microscope (TEM) images of irradiated olivine show the presence of npFe$^{0}$.

\section{Results and discussions}
\label{sect:Results and discussions}

\subsection{Different continua on spectral deconvolution}
\label{subsec:Different continua}

According to the definition of the MGM, the Gaussian fitting of the spectral curve needs to be processed by the continuum removal method. Flat line continuum (Method F) (Fig. 4) is the original choice by \cite{Sunshine1998Determining}. This method should be used with caution for spectral deconvolution, due to the fact that such flatness spectra in the near-infrared (NIR) regions are almost nonexistent of the mafic rocks, mixed minerals or lunar regolith. Compared with the widely used continuum removal method developed by \cite{clark1984reflectance}, this method will significantly overstate the absorption strength of the slope spectra in the ultraviolet and visible regions, when the continuum is far from the curve of visible region.

Oblique line continuum corresponds to Method O, the tangent continua contact the two inflection regions near 700 nm and 1800 nm. Cases in Fig. 5 show that this method will significantly overestimate the NIR region ranged from 2.0 $\mu$m to 2.6 $\mu$m. Actually, there is no significant absorption feature in this region. \cite{isaacson2011remote} used a similar continuum removal method to study the lunar olivine-dominated spectral in an individual absorption region. It makes some unrealistic physical modeling to fit the spectra in the longer wavelength region ($>$2.0 $\mu$m).

\begin{figure}
	\centering
		\begin{minipage}{7cm}
			\centering
			\includegraphics[width=7.2cm]{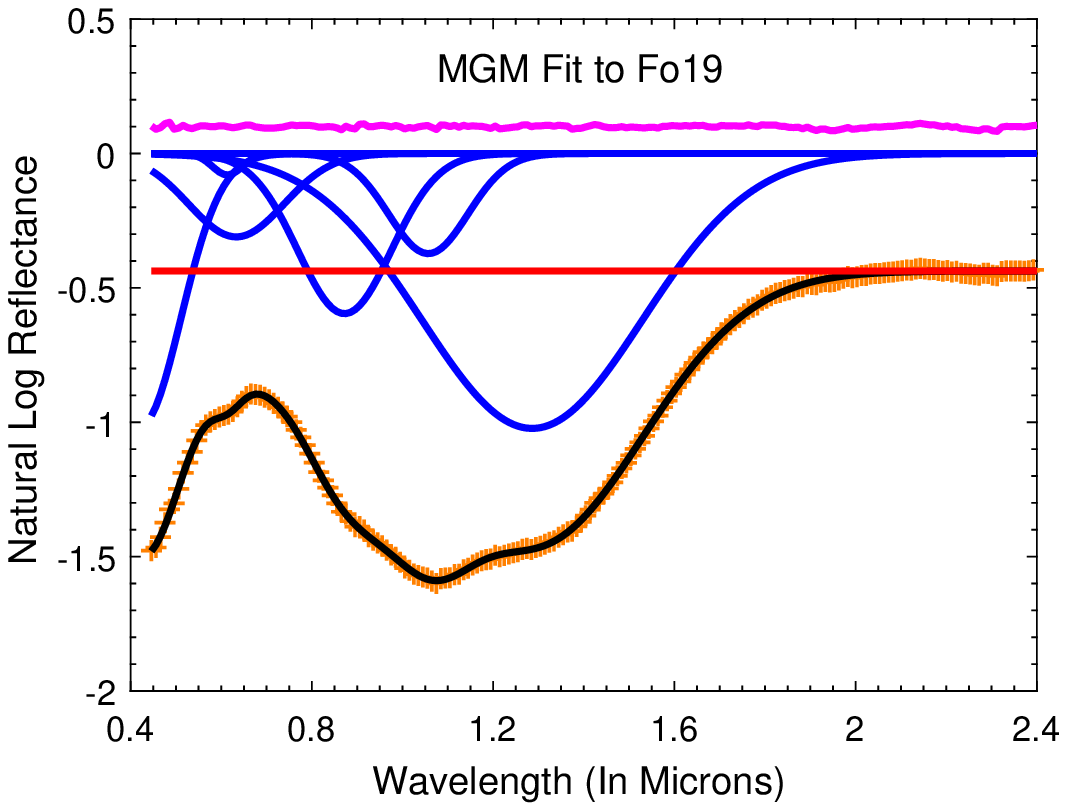}
		\end{minipage}
		\begin{minipage}{7cm}
			\centering
			\includegraphics[width=7.2cm]{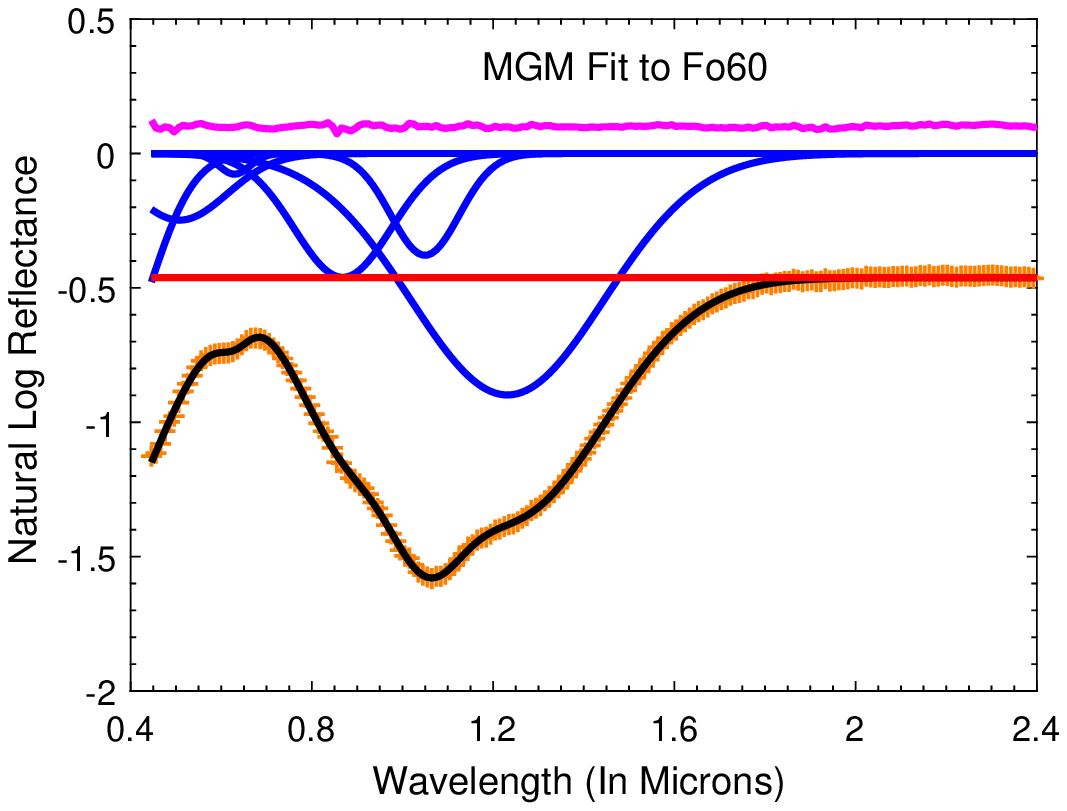}
		\end{minipage}
	\caption{Example of different Fo\# olivine spectral deconvolution results from flat line continuum removal method (Method F) of MGM}
	\label{Fig4}
\end{figure}

\begin{figure}
	\centering
		\begin{minipage}{7cm}
			\centering
			\includegraphics[width=7.2cm]{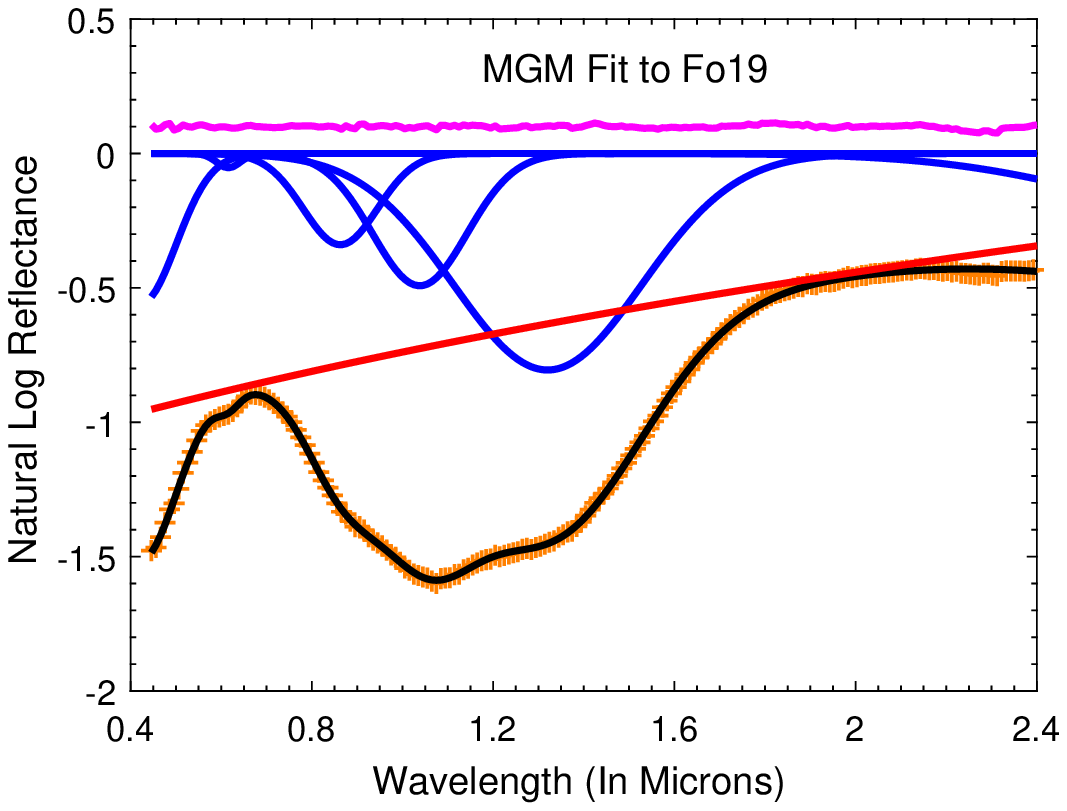}
		\end{minipage}
		\begin{minipage}{7cm}
			\centering
			\includegraphics[width=7.2cm]{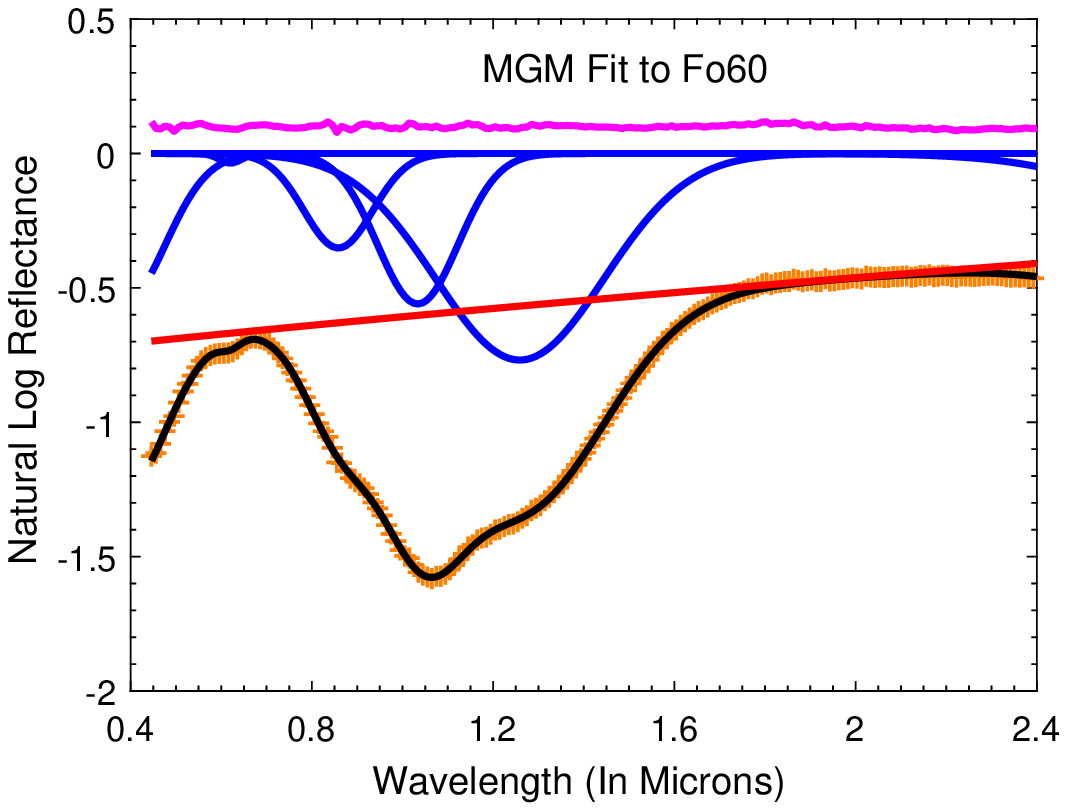}
		\end{minipage}
	\caption{Example of different Fo\# olivine spectral deconvolution results from oblique line continuum removal method (Method O) of MGM}
	\label{Fig5}
\end{figure}

Polynomial continuum corresponds to Method P, the logarithm of a second-order polynomial curve was used as the continuum to be removed in the MGM (as shown in Fig. 6). \cite{Sunshine1998Determining} used this method to fit the olivine-rich asteroid spectrum. However, they did not restrict the parameters of continuum so that a small range of variable Fo\# was accepted. With this method, the overall shape of the continuum can well describe the general spectral trend.

For Method E, the energy-wavelength polynomial continuum does not need to convert to natural logarithmic space, because it was originally defined in the natural logarithmic space (Fig. 7). This method is similar to a combination of Method O and Method P. Olivine diagnostic spectral feature (700 $\sim$ 1800 nm) can be well described, but sometimes it may underestimate the absorption in the visible region.

\begin{figure}
	\centering
		\begin{minipage}{7cm}
			\centering
			\includegraphics[width=7.2cm]{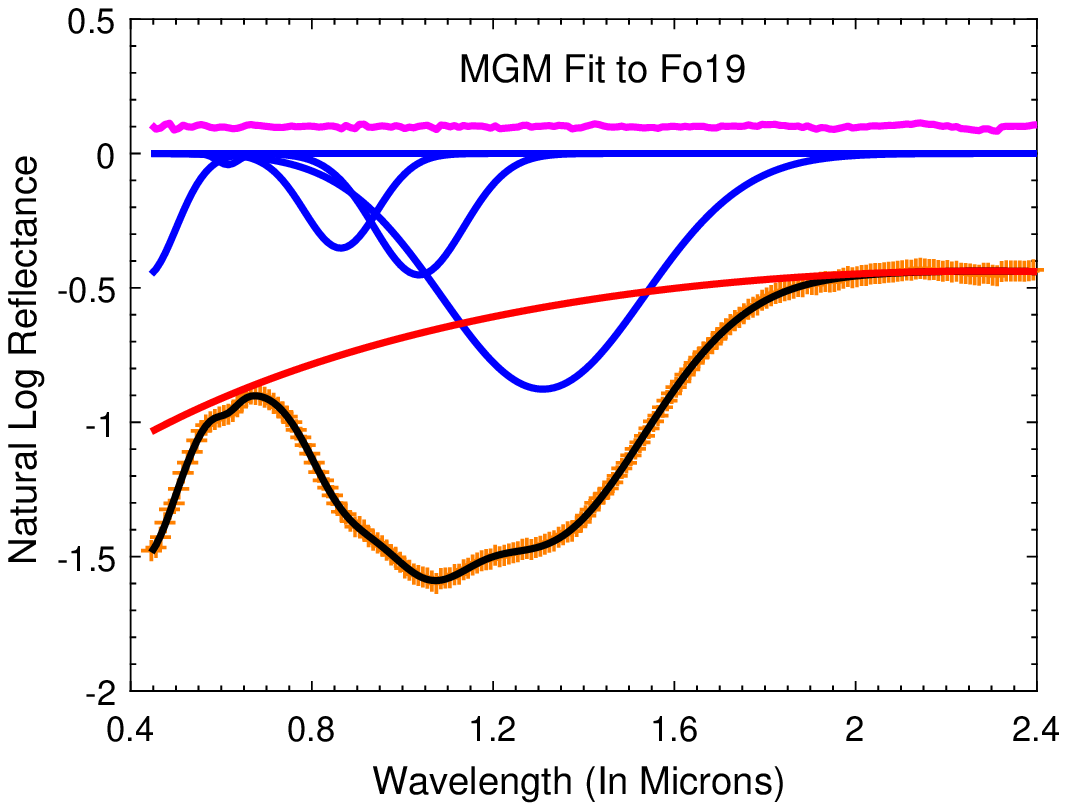}
		\end{minipage}
		\begin{minipage}{7cm}
			\centering
			\includegraphics[width=7.2cm]{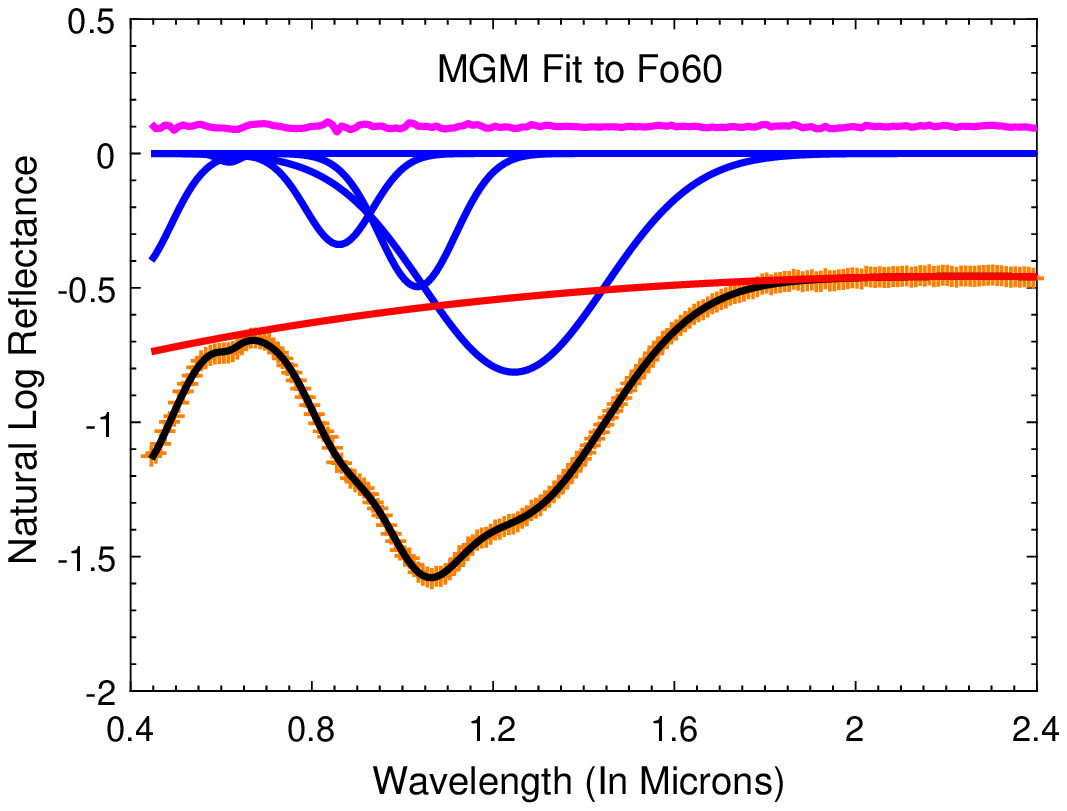}
		\end{minipage}
	\caption {Example of different Fo\# olivine spectral deconvolution results from second-order polynomial continuum removal method (Method P) of MGM}
	\label{Fig6}
\end{figure}

\begin{figure}
	\centering
		\begin{minipage}{7cm}
			\centering
			\includegraphics[width=7.2cm]{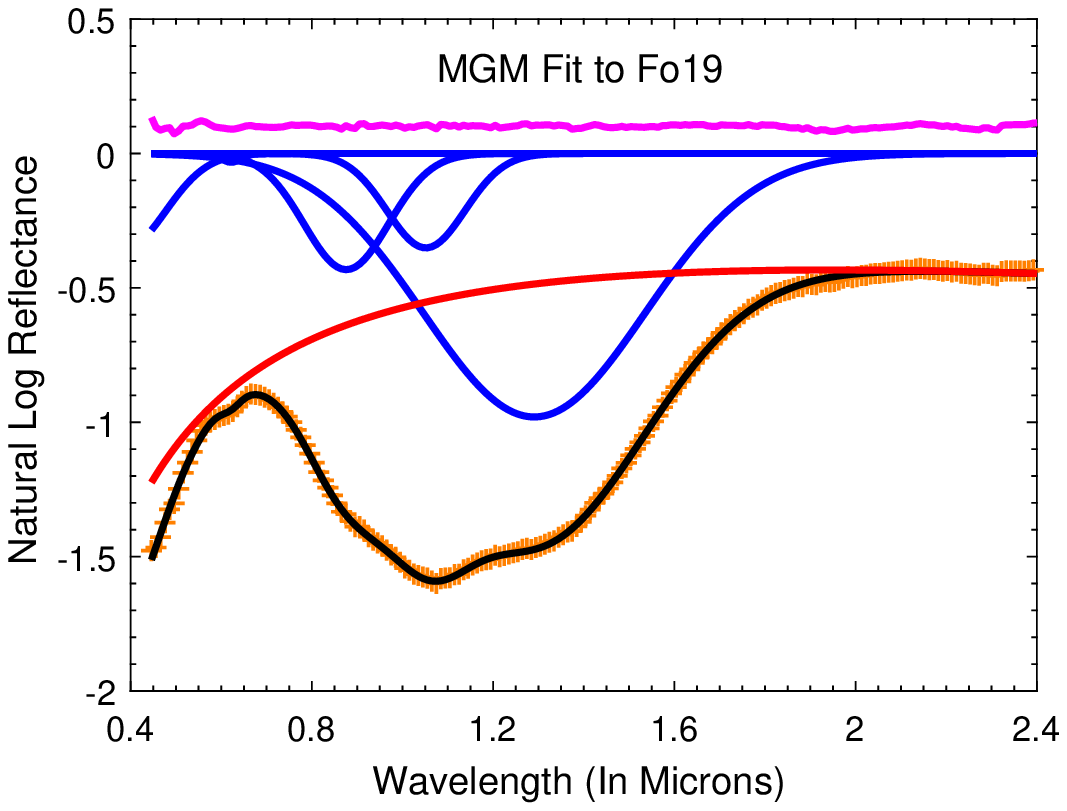}
		\end{minipage}
		\begin{minipage}{7cm}
			\centering
			\includegraphics[width=7.2cm]{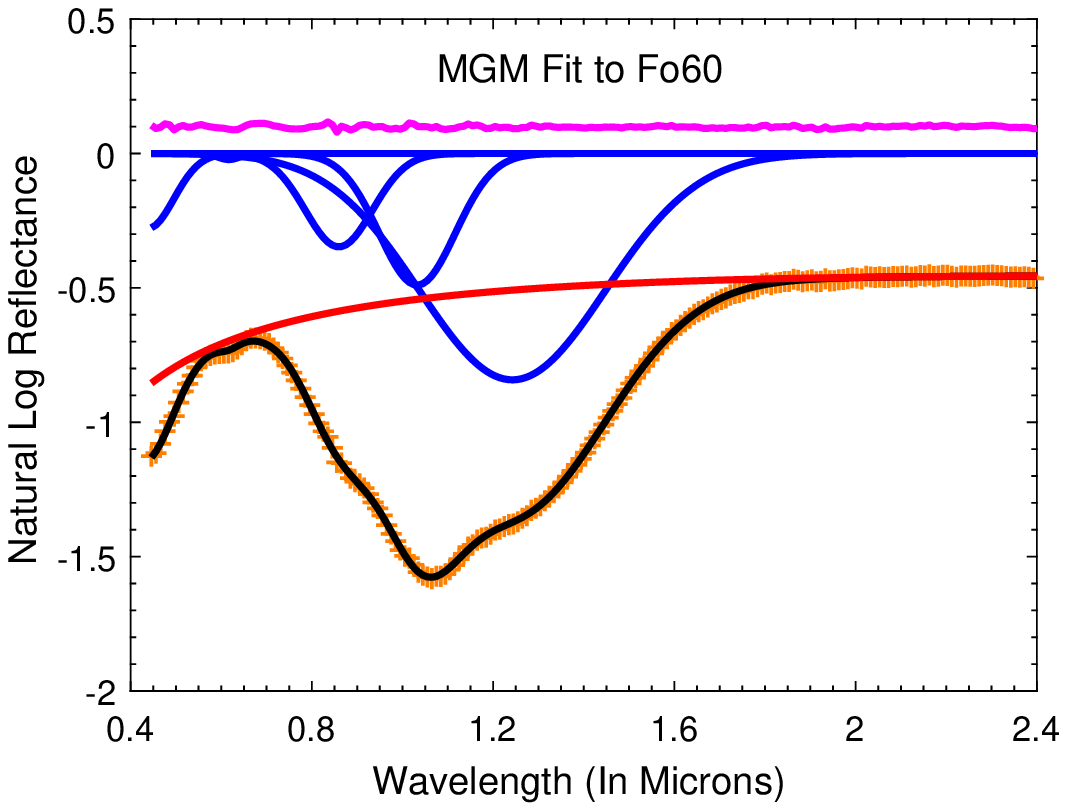}
		\end{minipage}
	\caption {Example of different Fo\# olivine spectral deconvolution results from energy-wavelength polynomial continuum removal method (Method E) of MGM}
	\label{Fig7}
\end{figure}

\subsection{Variations of the diagnostic band center}
\label{subsec:Variations}

In order to reveal the olivine composition from the MGM deconvolution results obtained using different continuum removal methods, it is essential to recalculate the three regression trend lines established by \cite{Sunshine1998Determining}. In addition, the position of the composite band center is also widely used to to determine the composition of olivine or pyroxene. Fig. 8 shows the variations of the olivine diagnostic band center position with their regression lines and the composite band centers. Table 2 shows the linear regression equations of the three diagnostic band centers, the composite band centers and the residuals. Comparison of the results of the four continuum removal methods and the trend lines of the composite band centers are shown in Fig.9.

\begin{figure}
	\centering
		\begin{minipage}{7cm}
			\centering
			\includegraphics[width=7.3cm]{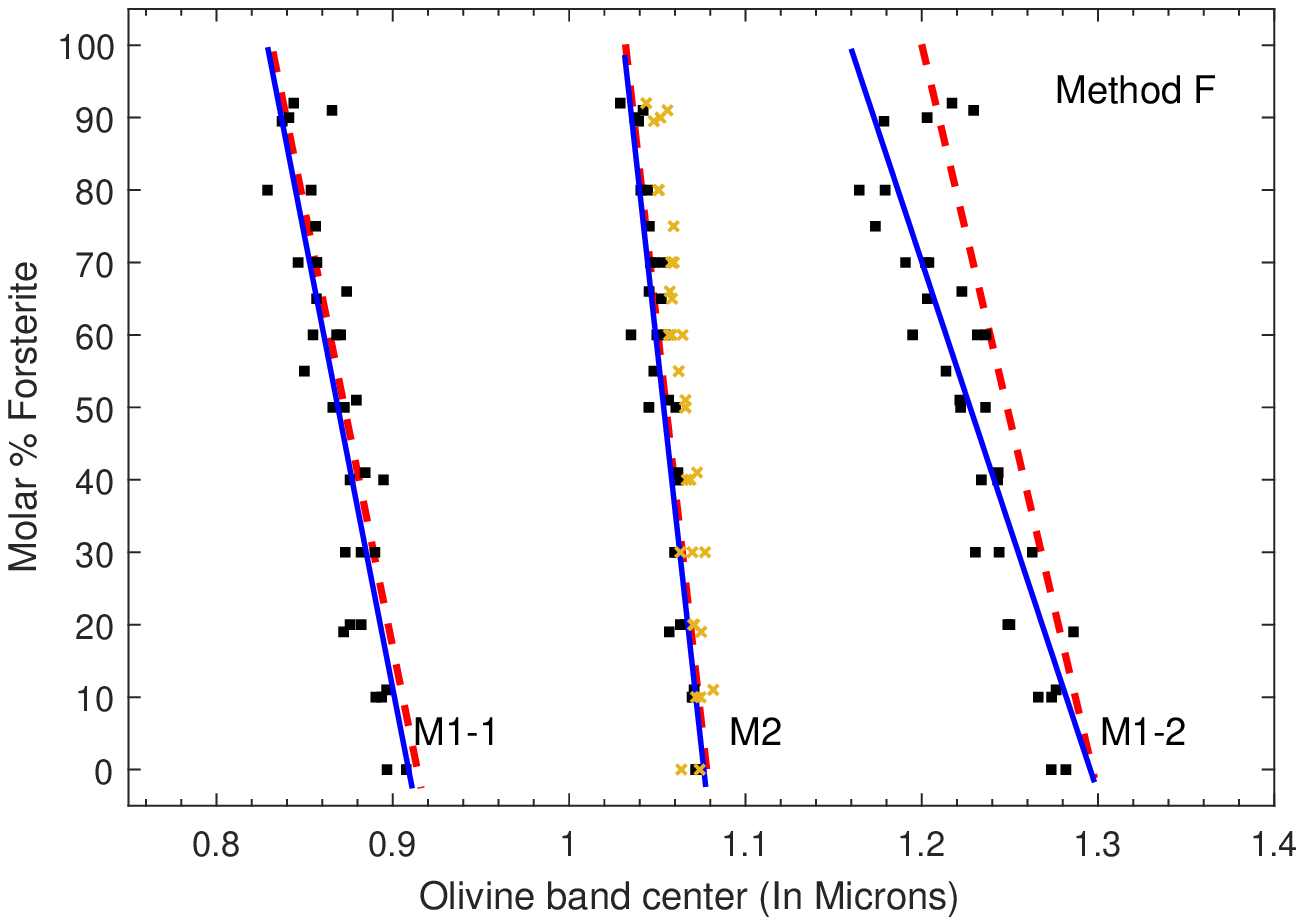}
		\end{minipage}
		\begin{minipage}{7cm}
			\centering
			\includegraphics[width=7.3cm]{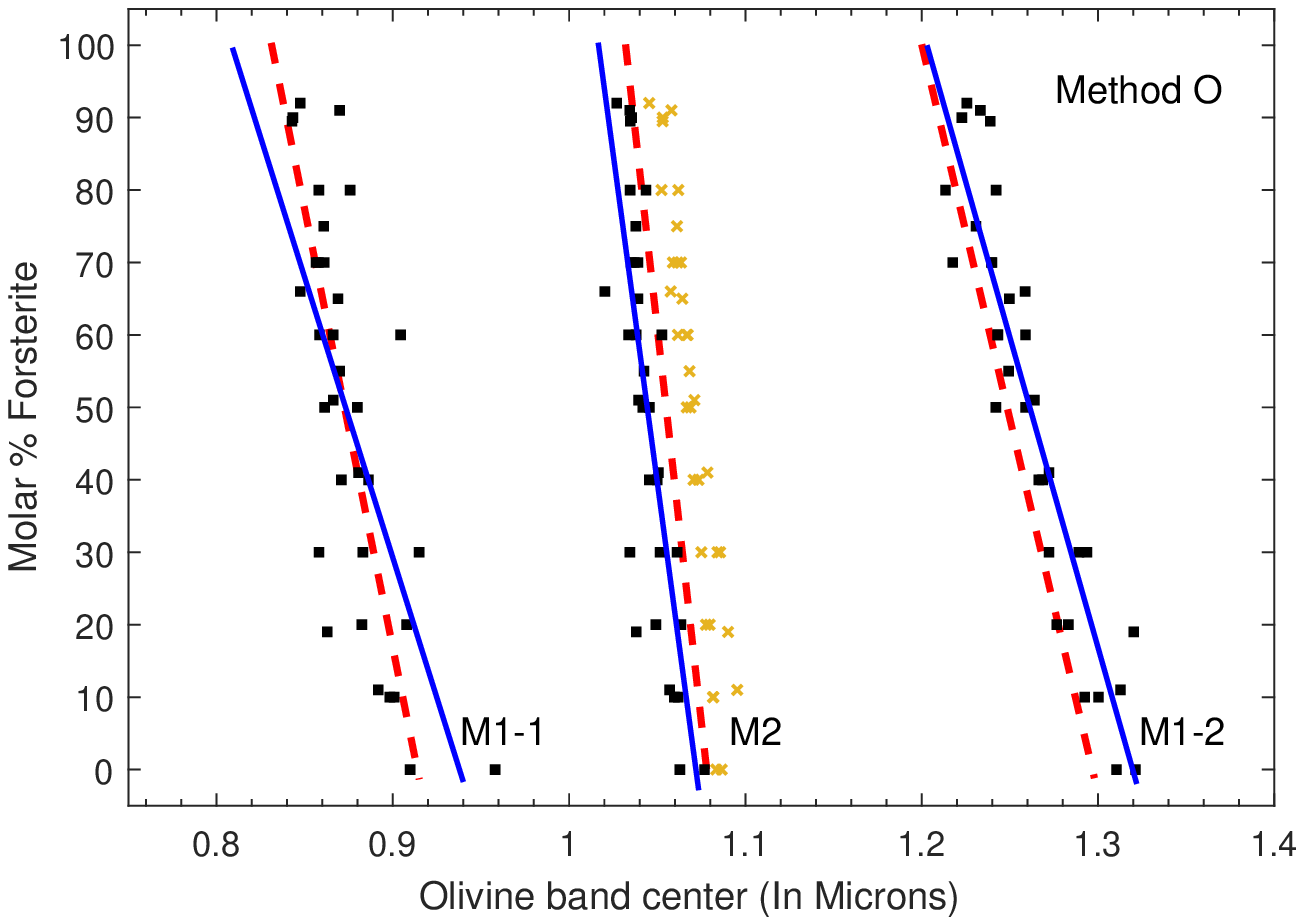}
		\end{minipage}
		\begin{minipage}{7cm}
			\centering
			\includegraphics[width=7.3cm]{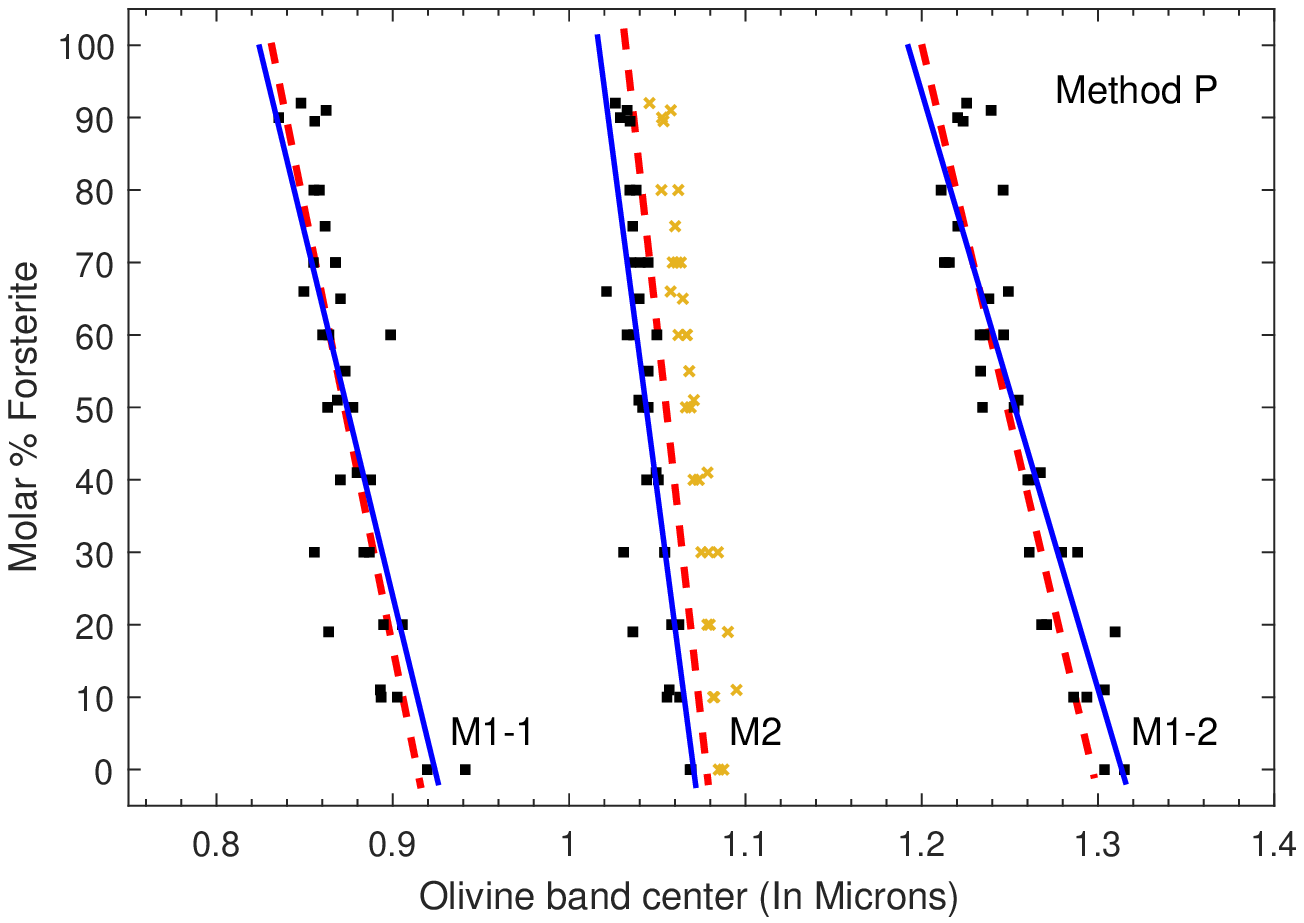}
		\end{minipage}
		\begin{minipage}{7cm}
			\centering
			\includegraphics[width=7.3cm]{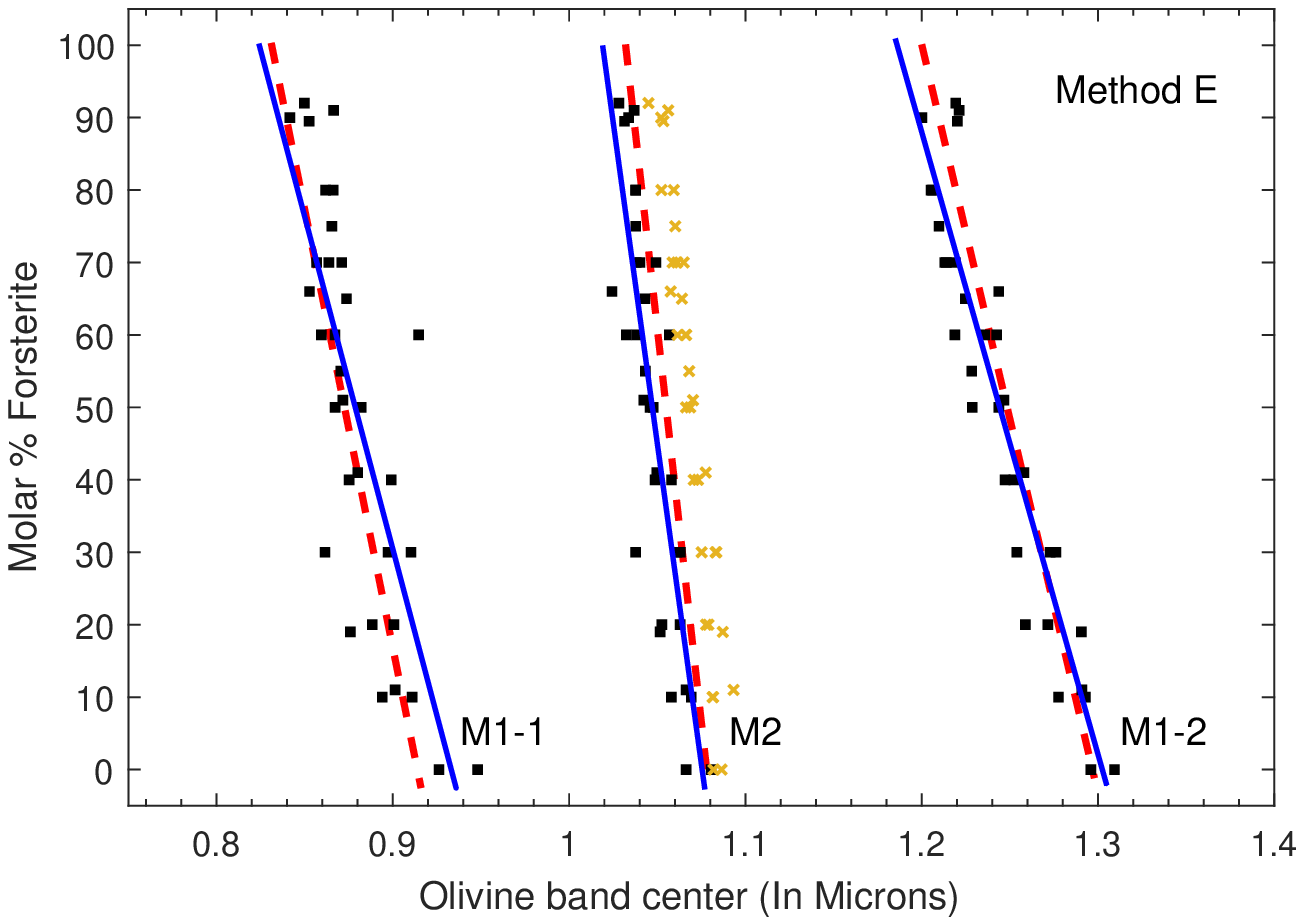}
		\end{minipage}
	\caption {Band centers of the olivine spectral deconvolution results from four continuum removal methods of MGM are compared with the trends defined by \cite{Sunshine1998Determining}. For this figure, black square symbols present the deconvolution results obtained in this paper, yellow cross present  the composite band centers, solid lines present these result trend lines, and red dashed lines present the trend lines defined by \cite{Sunshine1998Determining}.}
	\label{Fig8}
\end{figure}

\begin{figure}
	\centering
	\includegraphics[width=15.0cm, angle=0]{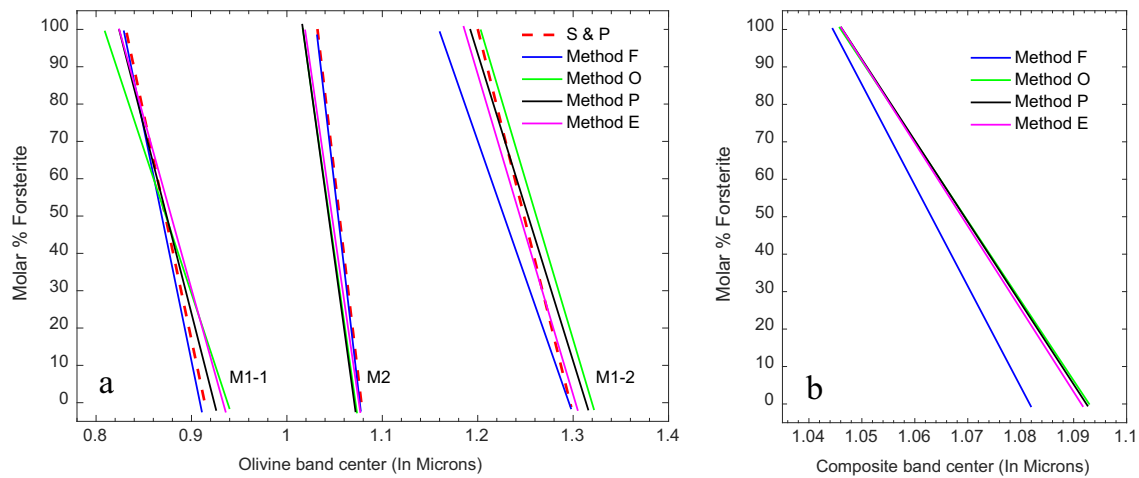}
	\caption{Comparison of the results of the four continuum removal methods and the trend lines of the composite band centers  trend lines of composite band centers obtained by four continuum removal methods. S\&P represents the trend lines defined by \cite{Sunshine1998Determining}.}
	\label{Fig9}
\end{figure}

As expected, with each method, olivine diagnostic band centers move toward longer wavelength with increasing iron content (decreasing Fo\#), and the two M1 band centers move faster than M2. The trend lines of M1-1 and M2 bands obtained by Method F are very close to that in \cite{Sunshine1998Determining}, but there are some changes in M1-2 band, it may due to that we used more sample data and/or used a different offset. In contrast to Method F, the linear regression equation system of the other three methods all show that the slope rates are less than the corresponding values in \cite{Sunshine1998Determining}. As discussed in Section 2.3 and 3.1, different continua in the MGM deconvolution will significantly influence the location and distribution of the band center, especially the trend of M1-1 and M2. With Method O, Method P and Method E, the trend lines of M2 band center position appear to be shifted by a few nanometers toward short wavelengths with a smaller slope than the results of \cite{Sunshine1998Determining}.

We note that the spectral curve becomes flat with increased Fo\#, especially when Fo\# $>$ 90 (Fig. 1). In addition, deconvolution results always shift a little bit toward longer wavelength when Fo\# $>$ 80. The absolute spectral strength of olivine will decrease and spectra curve will become flattened with decreasing iron content. Therefore, if the absorption is too weak, the interpretation of the deconvolution results should be taken cautiously.

\begin{table}
\bc
\begin{minipage}[]{140mm}
\caption[]{Linear regression equations of the three diagnostic band centers with different methods\label{tab2}}\end{minipage}
\setlength{\tabcolsep}{0.7 mm}
\small
 \begin{tabular}{cccccccccccccccc}
  \hline\noalign{\smallskip}
 Band& \multicolumn{3}{c}{M1-1} & &\multicolumn{3}{c}{M2} & &\multicolumn{3}{c}{M1-2} & &\multicolumn{3}{c}{Composite Band}\\
 	\cline{2-4} \cline{6-8} \cline{10-12} \cline{14-16}
  Method&	a&	b&	residual& &		a&	b&	residual& &		a&	b&	residual& &		a&	b&	residual\\
    \hline\noalign{\smallskip}
  F	&-1247.4&	1133.8&	79.52 & &	-2195.9	&2363.7	&64.44	& &	-733.42&	950.25&	80.47& &	-2691.8&	2911.7&	77.42\\
  O	&-773.65&	725.53&	103.97& &	-1810.1	&1940.3	&96.70	& &	-856.71&	1130.6&	59.06& &	-2131.8&	2329.9&	56.81\\
  P	&-1002.2&	925.89&	94.48 & &	-1857.4	&1988.6	&92.83 & &	-823.82&	1082.1&	68.71& &	-2170.9&	2371.5&	51.72\\
  E	&-918.19&	856.8&	96.18 & & 	-1770.7	&1904.3	&84.61	& &	-859.06 &    1118.9&54.70& &	-2215.4&	2418.0&	51.59\\
  	\noalign{\smallskip}\hline
 \end{tabular}
	\ec
	\tablecomments{1.0\textwidth}{The expression of the linear regression equation is: Fo\# = a*$\lambda$+b, where a and b are variable coefficients; $\lambda$ is wavelength in microns.}
\end{table}

For the MGM deconvolution of the continuum removal method, Fig. 8 and Table 2 show the linear regression equations and the residuals. Method F is the original choice by \cite{Sunshine1998Determining}. The trend lines of M1-1 and M2 in this paper are almost the same as those in \cite{Sunshine1998Determining}, while the centers of M1-2 are more dispersed, which indicates that this method may be not suitable for continuum removal.

Method O is a simple and feasible method, but compared with other methods, the sum of residuals of all three diagnostic band centers is the largest. Moreover, the oblique line continuum will significantly overestimate the NIR region longer than $\sim$2.0 $\mu$m. As a result, it needs to use some Gaussian curves without physical significance to fit the spectral curves. Fortunately, the diagnostic spectral characteristics of olivine does not exceed this region, hence we can restrict the spectral range to $\sim$0.6 $\mu$m $<$ wavelength $<$ $\sim$1.8 $\mu$m to study the olivine composition. However, we still do not recommend this approach due to its large residuals.

Method P is commonly used in the MGM deconvolution. With appropriate parameters, the overall shape of the smooth curved continuum can well describe the general spectral trend. Compared with other methods, Method P is also a simple and effective method. More importantly, this method does not need to initialize the Gaussian curves without physical significance. 

Method E was developed for deconvolving the reflectance spectra of lunar soils. It is not widely used in the MGM deconvolution, maybe because it requires complex and harsh initialization parameters. Furthermore, with the iterative parameters of this polynomial continuum, the stochastic non-linear inverse algorithm sometimes makes the new parameters worse than the old do. In this case, the deconvolution result is untrustworthy. This method requires manual reexamination to make sure the result is correct, which will bring more labor requirements. In view of these, Method E is not suitable for defining the continuum for plenty of remote sensing data.

In Fig. 9a, a comparison of the results obtained by the four continuum removal methods and the original method is clearly shown. What is more, in Fig. 9b, the trend lines of the composite band centers of Method O, P, and E are close, but the trend line of Method F is significantly different from others. Moreover, \cite{clark1984reflectance} pointed out that using a flat continuum to calculate the maximum absorption position of the reflectance spectrum is incorrect, which also indicates that the trend lines in \cite{Sunshine1998Determining} should be used with caution. Clearly, in terms of operability and practicality, Method P, the logarithm of a second-order polynomial continuum maybe the best choice in this paper for further research.

\subsection{Comparison between Method P and Sunshine \& Pieters (1998)}
\label{subsec:Comparison}

It is worth noting that the trend lines of the diagnostic band centers are some different with it defined by \cite{Sunshine1998Determining} and Method P. The band center and deviation obtained by these two methods with Fo\# changing in order are shown in Table 3. The deviations indicate that the corresponding band centers are quite different between the deconvolution results of Method P and \cite{Sunshine1998Determining}. Evidently, the deviation on M1-1 band is more than -5 $\sim$ 8 nm, even the band centers with a well linear trend, the deviation on M2 and M1-2 are more than -14 nm and -5 $\sim$ 14 nm respectively. Such a wide shift of the band center will require careful handling of deconvolution results.

\begin{table}
	\bc
	\begin{minipage}[]{150mm}
		\caption[]{Linear regression equations of the three diagnostic band centers with Method S\&P and Method P\label{tab3}}\end{minipage}
	\setlength{\tabcolsep}{2.0 mm}
	\small
	\begin{tabular}{cccccccccccc}
		\hline\noalign{\smallskip}
		\multirow{2}*{Fo\#}&\multicolumn{3}{c}{M1-1 (nm)}& &\multicolumn{3}{c}{M2 (nm)}&&\multicolumn{3}{c}{M1-2 (nm)}\\
		\cline{2-4} \cline{6-8} \cline{10-12}
		&S\&P&Method P&Deviation&&S\&P&Method P&Deviation&&S\&P&Method P&Deviation\\
		\hline\noalign{\smallskip}
		10&	905.6&	913.9&	8.3	& &	1073.4&	1065.3&	-8.2 & &	1287.2&	1301.4&	14.2\\
		20&	897.4&	903.9&	6.5	& &	1068.8&	1059.9&	-9.0 & &	1277.5&	1289.2&	11.7\\
		30&	889.1&	893.9&	4.8	& &	1064.2&	1054.5&	-9.7 & &	1267.8&	1277.1&	9.3\\
		40&	880.8&	883.9&	3.1	& &	1059.6&	1049.1&	-10.5& &	1258.2&	1265.0&	6.8\\
		50&	872.6&	874.0&	1.4	& &	1055.0&	1043.7&	-11.3& &	1248.5&	1252.8&	4.3\\
		60&	864.3&	864.0&	-0.3& &	1050.4&	1038.3&	-12.1& &	1238.8&	1240.7&	1.9\\
		70&	856.0&	854.0&	-2.0& &	1045.8&	1032.9&	-12.9& &	1229.1&	1228.5&	-0.6\\
		80&	847.8&	844.0&	-3.7& &	1041.2&	1027.6&	-13.7& &	1219.4&	1216.4&	-3.0\\
		90&	839.5&	834.1&	-5.5& &	1036.6&	1022.2&	-14.5& &	1209.8&	1204.3&	-5.5\\
		\noalign{\smallskip}\hline
	\end{tabular}
	\ec
	\tablecomments{1.0\textwidth}{S\&P represents the trend lines in \cite{Sunshine1998Determining}. Deviation is equal to the result of Method P minus S\&P.}
\end{table}

\begin{table}
	\bc
	\begin{minipage}[]{150mm}
		\caption[]{ Different estimated results (Fo\#) and corrections between the two regression equations defined by Method S\&P and Method P \label{tab4}}\end{minipage}
	\setlength{\tabcolsep}{1.0 mm}
	\small
	\begin{tabular}{cccccccccccc}
		\hline\noalign{\smallskip}
		Band& \multicolumn{3}{c}{Fo\# (M1-1)} & Band&\multicolumn{3}{c}{Fo\# (M2)} & Band& \multicolumn{3}{c}{Fo\# (M1-2)} \\
		\cline{2-4} \cline{6-8} \cline{10-12}
		Center&S\&P&Method P&Correctionn&Center&S\&P&Method P&Correctionn &Center&S\&P&Method P&Correctionn\\
		\hline\noalign{\smallskip}
		0.830&	101.5&	94.1&	-7.5&	1.030&	104.5&	75.5&	-29.0&	1.210&	89.8&	85.3&	-4.5\\
		0.840&	89.4&	84.0&	-5.4&	1.035&	93.6&	66.2&	-27.4&	1.220&	79.4&	77.0&	-2.4\\
		0.850&	77.3&	74.0&	-3.3&	1.040&	82.7&	56.9&	-25.8&	1.230&	69.1&	68.8&	-0.3\\
		0.860&	65.2&	64.0&	-1.2&	1.045&	71.8&	47.6&	-24.2&	1.240&	58.8&	60.6&	1.8\\
		0.870&	53.1&	54.0&	0.9	&	1.050&	60.9&	38.3&	-22.6&	1.250&	48.4&	52.3&	3.9\\
		0.880&	41.0&	44.0&	2.9	&	1.055&	50.1&	29.0&	-21.0&	1.260&	38.1&	44.1&	6.0\\
		0.890&	28.9&	33.9&	5.0	&	1.060&	39.2&	19.8&	-19.4&	1.270&	27.8&	35.8&	8.1\\
		0.900&	16.8&	23.9&	7.1	&	1.065&	28.3&	10.5&	-17.9&	1.280&	17.4&	27.6&	10.2\\
		0.910&	4.7	&	13.9&	9.2	&	1.070&	17.4&	1.2	&	-16.3&	1.290&	7.1	&	19.4&	12.3\\
		0.920&	-7.4&	3.9	&	11.3&	1.075&	6.6	&	-8.1&	-14.7&	1.300&	-3.3&	11.1&	14.4\\
		\noalign{\smallskip}\hline
	\end{tabular}
	\ec
	\tablecomments{0.97\textwidth}{Olivine band centers are in microns. S\&P represents the trend lines in \cite{Sunshine1998Determining}. Olivine with a Fo\# greater than 100 or a negative number does not exist.}
\end{table}

In fact, Method P, the logarithm of a second-order polynomial continuum, well matches the overall shape of the spectrum, is a popular choice now. In order to study the difference of the two different trend lines and estimating results, we also calculated the Fo\# and correction which obtained from the two regression equations defined by Method P and \cite{Sunshine1998Determining} while the band centers were altered in order. Table 4 shows that the estimated results of the two different trend lines are quite different. It will overestimate more than 7 mol\% for forsterite and underestimate more than 9 mol\% for fayalite on M1-1 band if we use the regression equations defined by \cite{Sunshine1998Determining}. Similarly, it will overestimate more than 4 mol\% for forsterite and underestimate more than 14 mol\% for fayalite on M1-2 band. Considering that the estimation of mafic mineral abundance and chemical composition difference is between 5\% $\sim$ 10\%, the difference between the results of these two methods is acceptable. Conversely, as for M2 band, the prediction results of these two methods vary greatly, and it will overestimate the Fo\# by about 15 to 30 from fayalite to forsterite. There is a need to determine which method is better.

\subsection{Validation and application: analysis of olivine-dominated meteorite spectra}
\label{subsec:Validation and application}

This research uses olivine-dominated meteorite spectra to test Method P and its trend lines described above. Brachinites are unshocked equigranular igneous material, which belongs to differentiated ultramafic achondrites but not a part of primitive achondrites (\citealt{nehru1983brachina, nehru1992brachinites, mittlefehldt2003brachinites}), and arguably may have come from differentiated asteroids (\citealt{mittlefehldt2003brachinites}).  Based on their similarities in reflectance spectra, the A-type asteroids are believed to be olivine-dominated and the parent bodies of brachinites (\citealt{nesvorny2009asteroidal}). In this work, the spectra of meteorite Brachinite EET99402 with different grain sizes (RELAB sample ID: MT-JMS-088, MT-JMS-329, TB-TJM-058) were chosen as the examples to study its chemical composition. The average mineral compositions, major, minor, and trace element contents of EET 99402 were analyzed by \cite{mittlefehldt2003brachinites}, they reported that when converted to weight percent, there are olivine 88.0\%, high-Ca pyroxene 4.7\%, plagioclase 6.1\%, spinel 1.0\%, and troilite 0.1\%. In addition, olivine has an average composition of Fo64.2.

\begin{table}
	\bc
	\begin{minipage}[]{150mm}
		\caption[]{MGM modelling parameters of the meteorite Brachinite EET99402 spectra\label{tab5}}\end{minipage}
	\setlength{\tabcolsep}{1.8 mm}
	\small
	\begin{tabular}{ccc|ccc|ccc|c}
		\hline\noalign{\smallskip}
		\multicolumn{2}{c}{Sample ID} &\multicolumn{2}{c}{MT-JMS-088}& &\multicolumn{2}{c}{MT-JMS-329}& &\multicolumn{2}{c}{TB-TJM-058}\\
		\hline\noalign{\smallskip}
		\multicolumn{2}{c}{Grain size} &\multicolumn{2}{c}{fine powder}& &\multicolumn{2}{c}{ $<$45 $\mu$m}& &\multicolumn{2}{c}{ $<$125 $\mu$m}\\
		\hline\noalign{\smallskip}
		\multicolumn{2}{c}{\multirow{2}*{Gaussian fitting parameter} }&Starting	&Final	& &	Starting&	Final& &Starting	&Final \\
		&& parameters	& parameters& &parameters& parameters& &parameters& parameters \\
		\hline\noalign{\smallskip}
				&parameter 1 	&3.50E-01& 	3.37E-01& & 	3.30E-01& 	3.54E-01& & 1.40E-01& 	1.38E-01\\
		Continuum& parameter 2	&5.00E-05&	5.14E-05& &		5.00E-05&	1.97E-05& &	5.00E-05&	4.51E-05\\
				&parameter 3	&-2.00E-08&	2.08E-08& &		-1.50E-08&	6.06E-09& &	-1.50E-08&	-1.29E-08\\
				&	&	&	&	&	&	&	&	&	\\
				&center	&	300&	292.6& & 	300&	295.2& & 	300&	323.3\\
		Band 1	&FWHM	&	160&	110.8& & 	100&	109.0& & 	160&	184.7\\
				&strength&	-1.0&	-1.21& &	-1.0&	-1.36& &	-1.0&	-1.04\\
				&	&	&	&	&	&	&	&	&	\\
				&center	&	400&	349.6& &	400&	374.7& &	480&	480.4\\
		Band 2	&FWHM	&	200&	250.9& &	100&	222.6& &	100&	97.5\\
				&strength&	-0.3&	-0.51& &	-0.3&	-0.44& &	-0.4&	-0.26\\
				&	&	&	&	&	&	&	&	&	\\
				&center	&	620&	653.5& &	620&	628.9& &	620&	621.5\\
		Band 3	&FWHM	&	70&		75.3& &	 	70&		49.2& &		70&		63.7\\
				&strength&	-0.02&	-0.0198& &	-0.02&	-0.0188& &	-0.02&	-0.0198\\
				&	&	&	&	&	&	&	&	&\\
				&center	&	870&	863.7& &	870&	863.1& &	870&	862.3\\
		M1-1	&FWHM	&	180&	183.7& &	180&	168.8& &	180&	173.8\\
				&strength&	-0.1&	-0.12& &	-0.1&	-0.11& &	-0.2&	-0.24\\
				&	&	&	&	&	&	&	&	&	\\
				&center	&	1050&	1041.4& &	1050&	1039.4& &	1050&	1036.1\\
		M2		&FWHM	&	200&	184.0& &	200&	187.9& &	200&	198.4\\
				&strength&	-0.25&	-0.23& &	-0.25&	-0.21& &	-0.35&	-0.41\\
				&	&	&	&	&	&	&	&	&\\	
				&center	&	1250&	1232.8& &	1250&	1225.2& &	1250&	1238.4\\
		M1-2	&FWHM	&	400&	384.8& &	400&	412.2& &	350&	401.1\\
				&strength&	-0.25&	-0.24& &	-0.22&	-0.24& &	-0.5&	-0.50\\
		\hline\noalign{\smallskip}
		\multicolumn{2}{c}{Root Mean Square Error} &\multicolumn{2}{c}{3.88E-03}& &\multicolumn{2}{c}{3.23E-03}& &\multicolumn{2}{c}{5.55E-03}\\
		\hline\noalign{\smallskip}
	\end{tabular}
	\ec
	\tablecomments{0.97\textwidth}{Band center and FWHM are in microns. Final parameters shown in this table are approximations values rather than exact values.}
\end{table}

The spectra of EET 99402 were deconvolved by MGM with continuum removal Method P, which aims to determine whether the prediction result of olivine compositions are consistent with their known chemistry. Fig. 10 shows the deconvolution results, while Table 5 lists the starting parameters used to initialize the spectra and final parameters obtained by MGM deconvolution. Furthermore, Fig. 10d. Shows the predicted results of the three spectra of ETT 99402. The predicted Fo\# of the three spectra are Fo60.3, Fo63.8, and Fo62.6, respectively, and the average value is Fo62.3. The estimated result is very close to its chemical component (Fo64.2), and it is superior to $\sim$Fo60 which calculated by \cite{Sunshine2007Olivine}. Probably because we used more samples with different grain sizes to make the estimates more accurate.

\begin{figure}
	\centering
		\begin{minipage}{7cm}
			\centering
			\includegraphics[width=7.4cm]{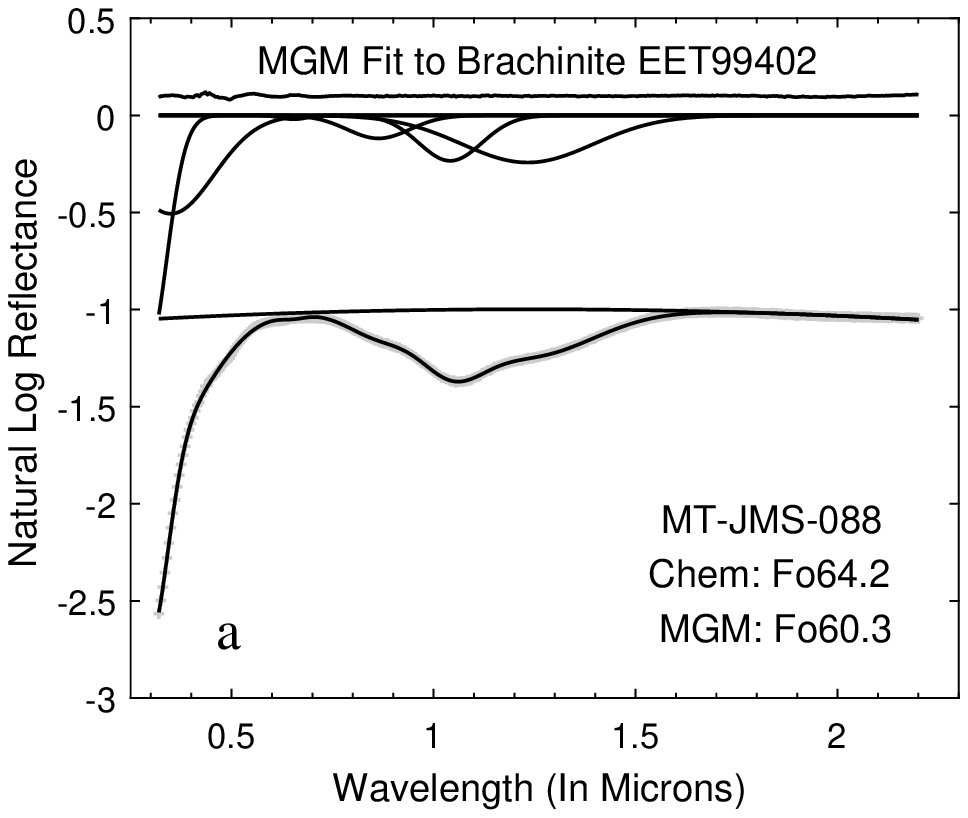}
		\end{minipage}
		\begin{minipage}{7cm}
			\centering
			\includegraphics[width=7.4cm]{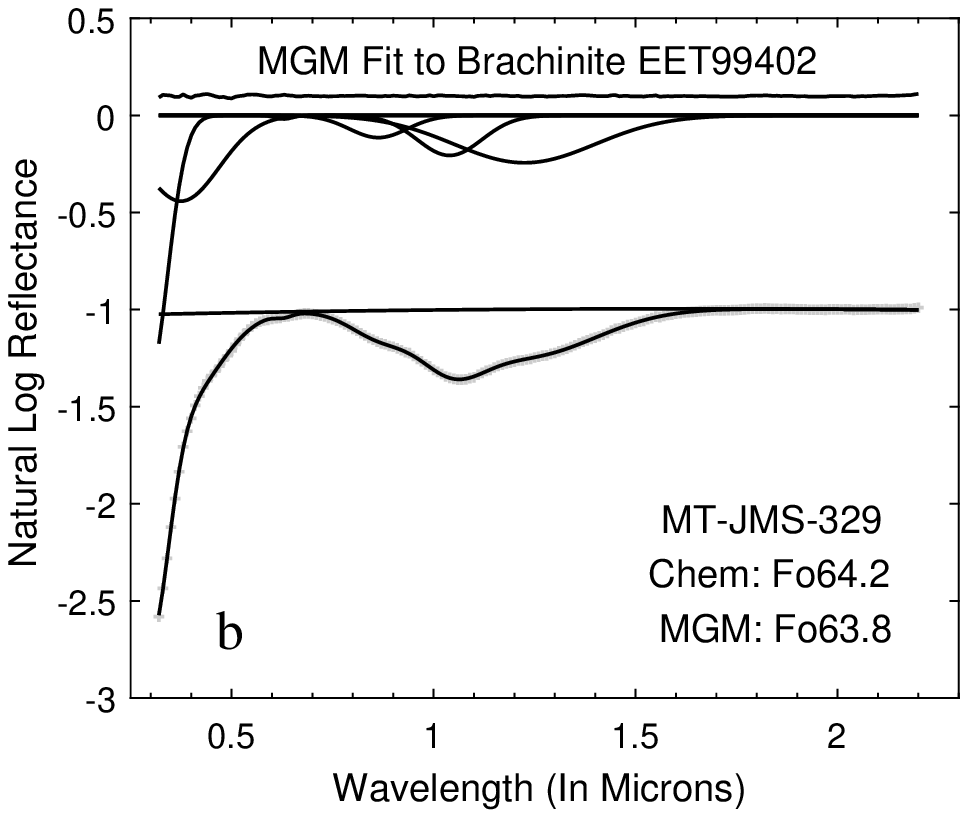}
		\end{minipage}
		\begin{minipage}{7cm}
			\centering
			\includegraphics[width=7.4cm]{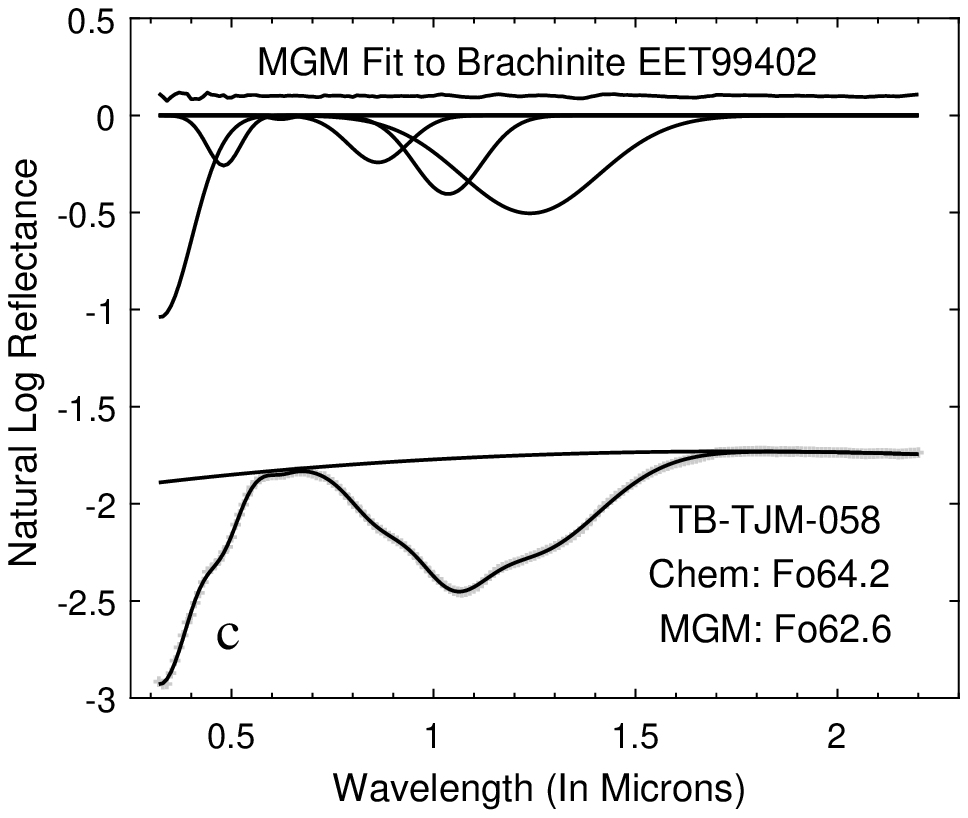}
		\end{minipage}
		\begin{minipage}{7cm}
			\centering
			\includegraphics[width=7.4cm]{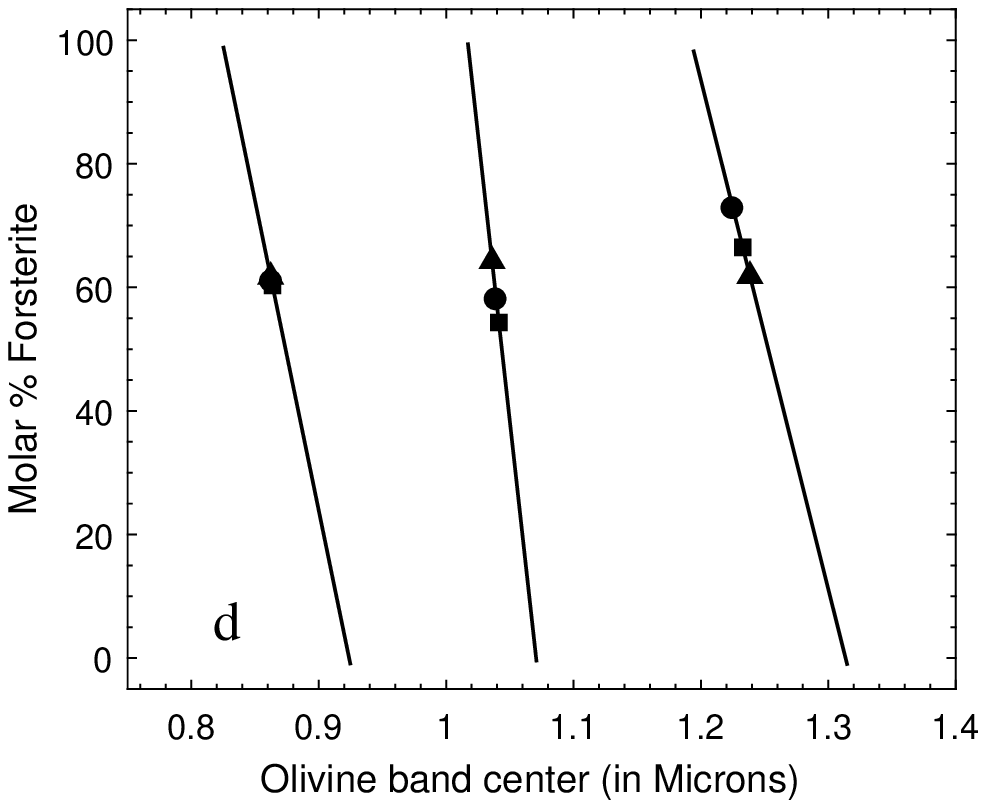}
		\end{minipage}
	\caption{MGM fit to meteorite Brachinite EET99402 with Method P and predict the mole fraction of Fo\#. Diagram a, b, and c are the MGM deconvolution results of Brachinite EET99402 spectra. Diagram d: solid lines are the regression equations of Method P; black squares, circles, and triangles represent the predicted Fo\# of MT-JMS-088, MT-JMS-329, and TB-TJM-058, respectively.}
	\label{Fig10}
\end{figure}

According to the comparison and discussion mentioned above, it should be known that if we use different continuum removal methods to deconvolve the spectrum with MGM, it will give significantly different absorption band center positions. These band centers are crucial for studying the olivine composition (Fo\#). Therefore, it should take caution to use the regression equation obtained by unknown or different continuum removal methods.

Generally speaking, the optimal continuum removal method mainly depends on the overall shape of the spectral curve. For a single simple absorption band, we usually choose a simple flat or oblique tangent line as the continuum. The widely used cubic splines continuum in \cite{clark1984reflectance} is similar to the tangent line continuum removal method. Shift to the logarithmic space of the reflectance, the oblique line will bend to have a little curvature at the short wavelength region. The physical significance of the continuum is not clear yet, if we do not want to increase the absorption bands without physical significance and excessively decrease the band intensity, the logarithm of a second-order polynomial continuum, which can match the overall shape of the varied spectral curves, may be the best choice.

\subsection{Application of MGM in space weathering analysis}
\label{subsec:Application}

\subsubsection{Reflectance spectra of irradiated olivine}
\label{susubbsec:irradiated olivine}

Based on the optimized MGM method mentioned above for estimating olivine composition, the spectral characteristics of olivine subjected to space weathering simulated in the laboratory was studied. The VNIR reflectance spectra of the samples with varied laser irradiation degrees as well as the original one are shown in Fig. 11a, while the same spectra normalized at 700 nm are shown in Fig. 11b.

After irradiation, those diagnostic spectral features turned weaker and shallower. As the number of irradiation increases, absorption intensity and albedo of the visible region (we chose the reflectance value at 700 nm) decrease faster than the NIR region (we chose the reflectance value at 1800 nm), which behaviors redden the spectra. It is in accordance with the lunar-style space weathering (\citealt{gaffey2010space}). According to the methods described above, Method F and Method O do not apply to the reflectance spectra of irradiated olivine. While, as mentioned above, Method P and Method E can be used for deconvolving the reflectance spectra of lunar soils. So,  these two continua removal methods will be used to study the reflectance spectra of irradiated olivine.

\begin{figure}
	\centering
	\includegraphics[width=15.5cm, angle=0]{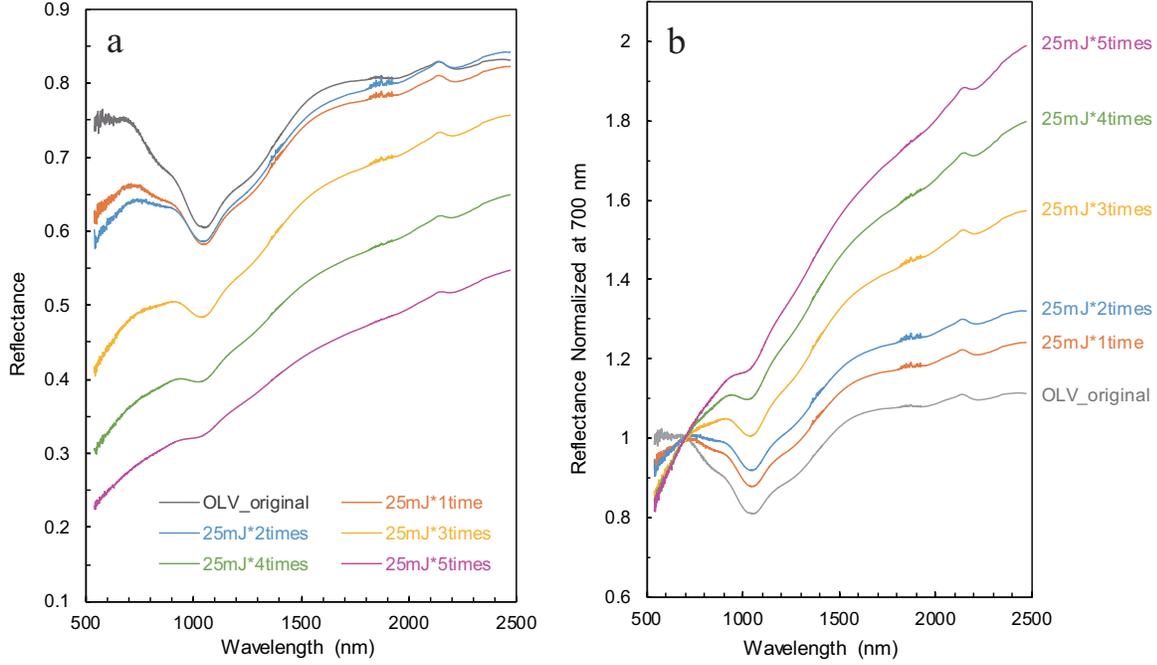}
	\caption{VNIR reflectance spectra and normalized spectra of olivine samples before and after irradiation. Diagram a: all spectra are measured relative to a Spectralon standard with 99\% nominal reflectance. Diagram b: the reflectance spectra of olivine samples normalized at 700 nm.}
	\label{Fig11}
\end{figure}

\subsubsection{Effects of space weathering on olivine diagnostic spectral features}
\label{susubbsec:Effects of space weathering}

The absorption center of olivine diagnostic spectra is near 1050 nm, and the edges of the wings are near 700 nm and 1800 nm as stated in Section 3.1. In order to quantify the variation of the olivine spectral curve in space weathering, we defined a function to characterize the stability of the spectral diagnostic features:

\begin{equation}
\label{Eqn6}
SSDF_{olivine}=\frac{R_{700}-R_{1050}}{R_{1800}-R_{1050}}
\end{equation}

where $SSDF_{olivine}$ is an abbreviation for stability of the spectral diagnostic features of olivine; $R_{700}$, $R_{1050}$, and $R_{1800}$ represent the reflectance or albedo at 700 nm, 1050 nm, and 1800 nm, respectively.

$SSDF_{olivine}$ does not depend on the absolute reflectance of the spectrum, relative reflectance spectrum is also applicable, especially the telescope spectra, which makes the Function (6) more accurate and practical when studying space weathering. Table 6 shows parameters characterizing the VNIR spectral curves of olivine before and after irradiations. As the number of radiation increases, the degree of spectral reddening increases, and the stability of spectral diagnostic features decreases.

\begin{table}
	\bc
	\begin{minipage}[]{150mm}
		\caption[]{Parameters characterizing the VNIR spectral curves of olivine before and after irradiation\label{tab6}}\end{minipage}
	\setlength{\tabcolsep}{3.0 mm}
	\small
	\begin{tabular}{cccccc}
		\hline\noalign{\smallskip}
		Olivine	&\multicolumn{3}{c}{Reflectance (or albedo)}&Reddening&SSDF\\
		\cline{2-4}
		samples 	&700 nm	&1050 nm&1800 nm&($R_{1800}$/$R_{700}$)& \\
		\hline\noalign{\smallskip}
		OLV-original	&0.7466	&0.6052	&0.8049	&1.0781	&0.7081\\
		25 mJ*1 time		&0.6626	&0.5825	&0.7772	&1.1730	&0.4114\\
		25 mJ*2 times		&0.6373	&0.5856	&0.7914	&1.2418	&0.2512\\
		25 mJ*3 times		&0.4810	&0.4844	&0.6972	&1.4495	&-0.0160\\
		25 mJ*4 times		&0.3612	&0.3988	&0.5727	&1.5855	&-0.2162\\
		25 mJ*5 times		&0.2752	&0.3254	&0.4716	&1.7137	&-0.3434\\
		\hline\noalign{\smallskip}
	\end{tabular}
	\ec
	\tablecomments{0.96\textwidth}{Function (6) shows the calculation method of SSDF as an abbreviation for the stability of the diagnostic features. When space weathering causes the olivine spectral slope to increase, the higher SSDF value, the clearer diagnostic features of the spectra and the lower degree of space weathering.}
\end{table}

To quantify the effects of pulsed laser irradiation on olivine spectra, these experiment spectra were firstly smoothed using the Savitzky-Golay (SG) filter procedure (\citealt{Savitzky1964Smoothing}). MGM with the logarithm of a second-order polynomial continuum removal method (Method P) and an Energy- wavelength polynomial continuum removal method (Method E) were used to deconvolve the olivine spectra measured above. However, Method E shows the instability of the algorithm and the complexity of parameter settings during the deconvolution process as described at the end of Section 3.2. So we cannot really recommend using this method to study space weathered spectra. But the data results obtained by this method are still shown below. In addition, we limit the VNIR feature discussions in the spectral region 500 $\sim$ 2000 nm, because the spurious peak near 2140 nm caused by Spectralon absorption can confuse and complicate mineral identifications (\citealt{Zhang2014Effects}). Fig. 12  shows the diagnostic spectral features of these irradiated olivine grains and the original one obtained by Method P, while Table 7 and Table 8 show the final MGM fitting parameters of these spectra obtained by Method P and Method E, respectively.

\begin{figure}
	\centering
		\begin{minipage}{7cm}
			\centering
			\includegraphics[width=7.4cm]{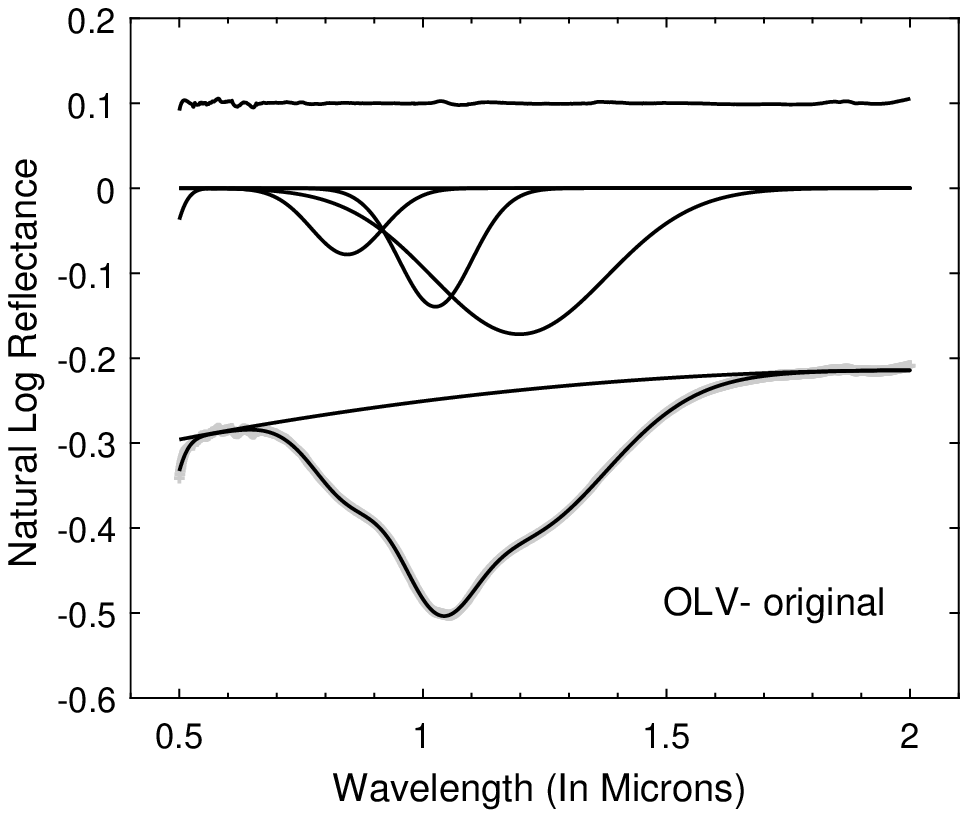}
		\end{minipage}
		\begin{minipage}{7cm}
			\centering
			\includegraphics[width=7.4cm]{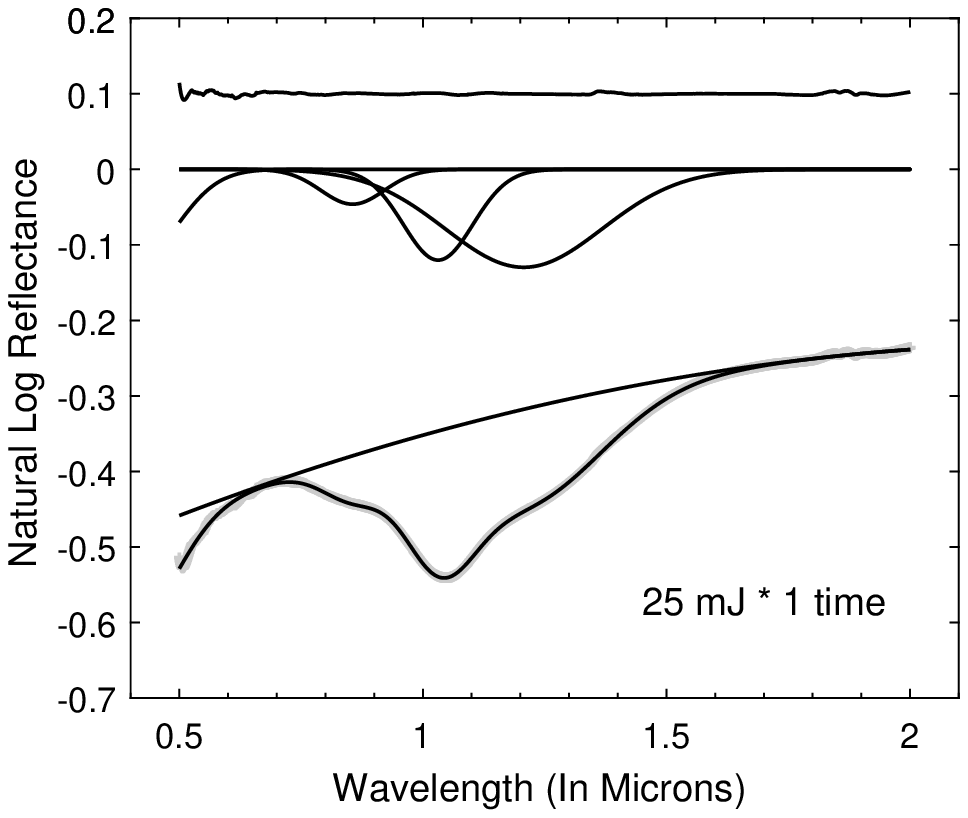}
		\end{minipage}
		\begin{minipage}{7cm}
			\centering
			\includegraphics[width=7.4cm]{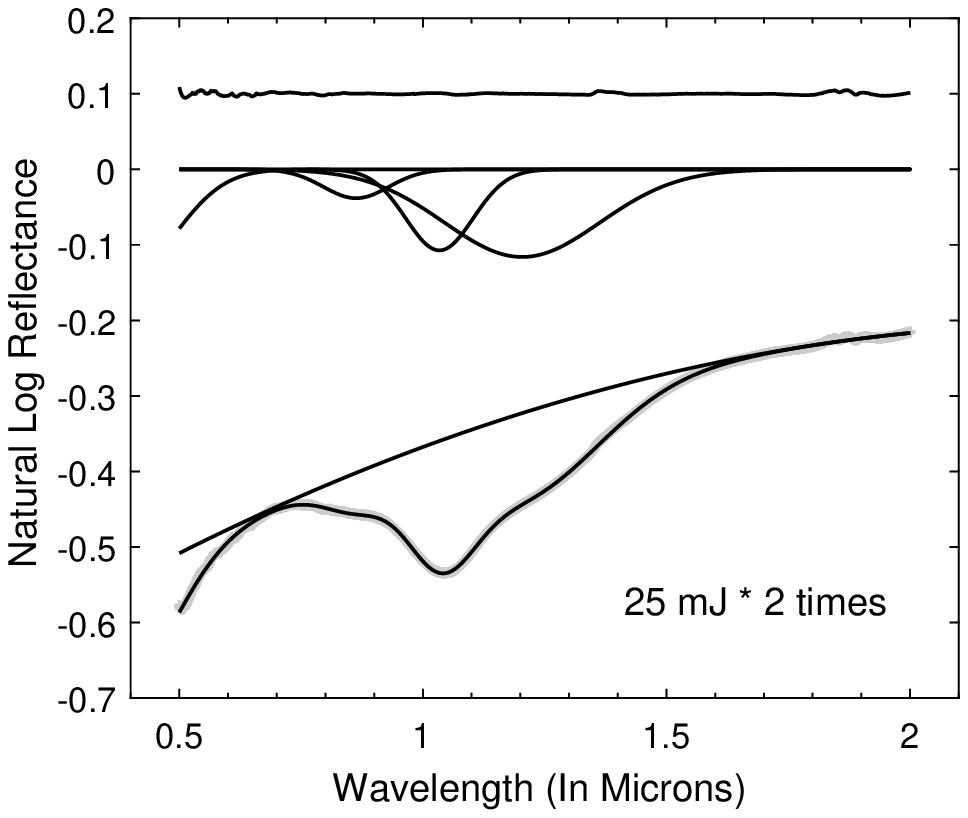}
		\end{minipage}
		\begin{minipage}{7cm}
			\centering
			\includegraphics[width=7.4cm]{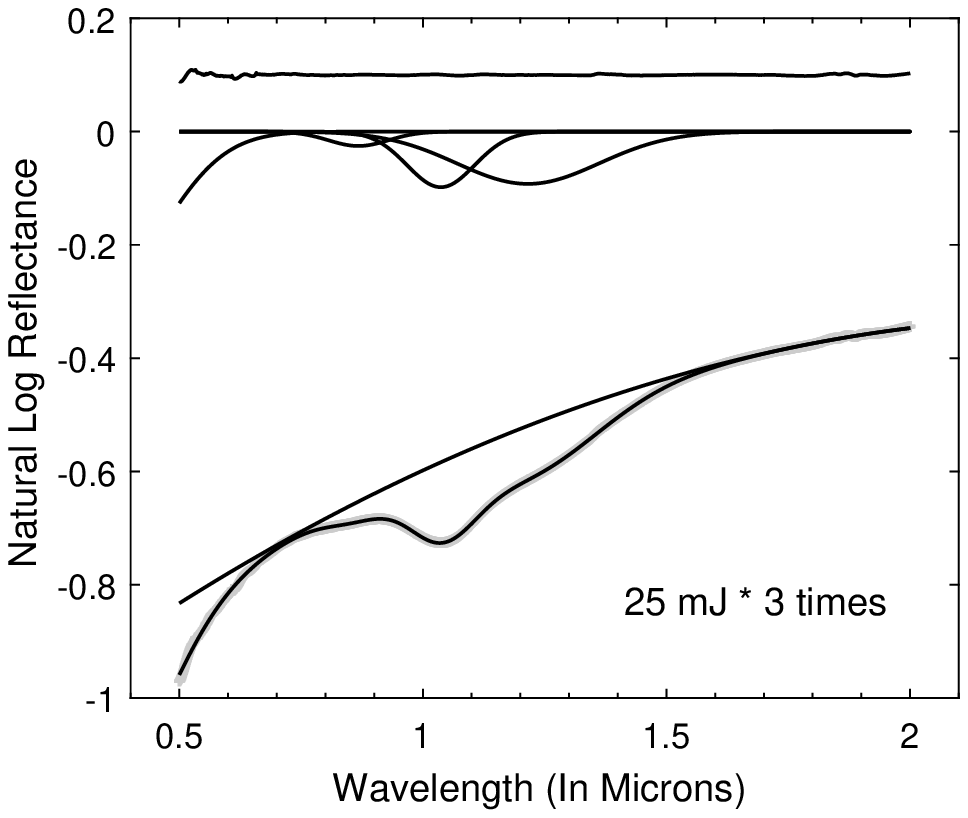}
		\end{minipage}
		\begin{minipage}{7cm}
			\centering
			\includegraphics[width=7.4cm]{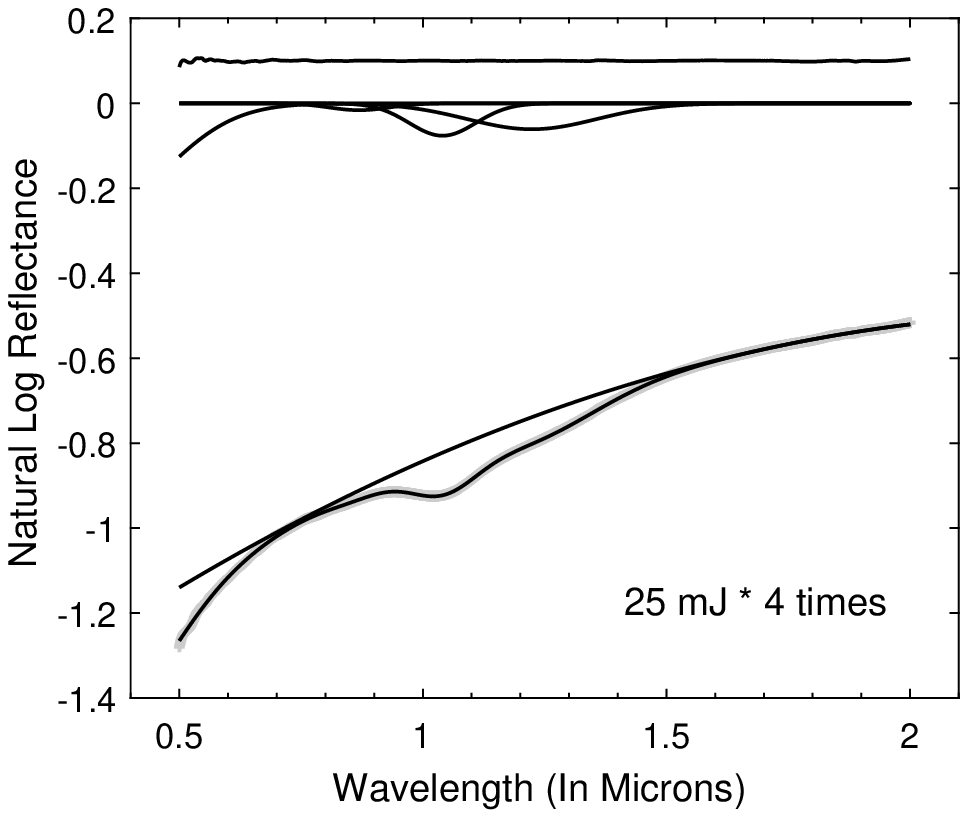}
		\end{minipage}
		\begin{minipage}{7cm}
			\centering
			\includegraphics[width=7.4cm]{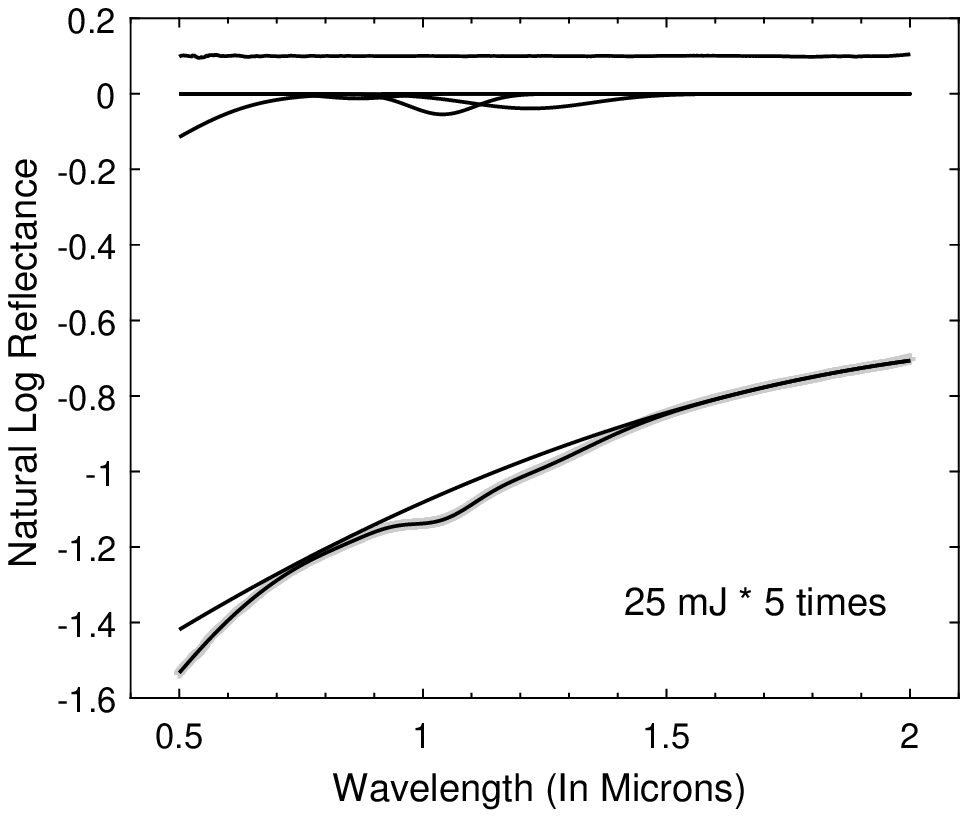}
		\end{minipage}
	\caption{MGM fit to the spectra of original fresh and varied degree irradiated olivine samples with Method P}
	\label{Fig12}
\end{figure}

\begin{table}
	\bc
	\begin{minipage}[]{150mm}
		\caption[]{Final MGM fitting parameters of the original fresh and irradiated olivine spectra with Method P\label{tab7}}\end{minipage}
	\setlength{\tabcolsep}{1.5 mm}
	\small
	\begin{tabular}{ccccccccccc}
		\hline\noalign{\smallskip}
		Olivine&\multicolumn{3}{c}{Band M1-1}&\multicolumn{3}{c}{Band M2}&\multicolumn{3}{c}{Band M1-2}&Estimated\\
		samples&center&FWHM&strength&center&FWHM&strength&center&FWHM&strength&Fo\#\\
		\hline\noalign{\smallskip}
		Original&844.6&173.5&-0.078&1025.9 &176.2&-0.139 &1197.9 &421.3&-0.172&85.9\\
		25mJ*1time&855.7&152.6&-0.046&1031.3&168.8 &-0.121 &1207.6&378.8 &-0.129&76.2\\
		25mJ*2times&862.9&158.0&-0.038&1033.9&162.4 &-0.107 &1203.8&376.3 &-0.116&73.2\\
		25mJ*3times&867.6&144.2&-0.025&1036.8&162.7 &-0.098 &1215.6&344.8 &-0.092&66.6\\
		25mJ*4times&869.3&145.1&-0.016&1040.3&161.2 &-0.076 &1223.8&314.1 &-0.061&61.6\\
		25mJ*5times&867.6&159.7&-0.012&1040.4&158.6 &-0.054 &1220.5&299.5 &-0.039&63.1\\
		\hline\noalign{\smallskip}
	\end{tabular}
	\ec
		\tablecomments{0.990\textwidth}{The estimated Fo\# result is the average of the predicted values of the three diagnostic bands.}
\end{table}

\begin{table}
	\bc
	\begin{minipage}[]{150mm}
		\caption[]{Final MGM fitting parameters of the original fresh and irradiated olivine spectra with Method E \label{tab8}}\end{minipage}
	\setlength{\tabcolsep}{1.5 mm}
	\small
	\begin{tabular}{ccccccccccc}
		\hline\noalign{\smallskip}
		Olivine&\multicolumn{3}{c}{Band M1-1}&\multicolumn{3}{c}{Band M2}&\multicolumn{3}{c}{Band M1-2}&Estimated\\
		samples&center&FWHM&strength&center&FWHM&strength&center&FWHM&strength&Fo\#\\
		\hline\noalign{\smallskip}
		Original &   847.7& 186.7& 	-0.086& 	1027.9& 175.2& -0.145& 	1201.6&405.7& 	-0.170& 83.1\\
		25mJ*1time&  855.1& 170.9&	-0.056& 	1032.0& 165.9& -0.120& 	1200.3&394.5& 	-0.142& 78.8\\
		25mJ*2times& 861.2& 181.9& 	-0.050& 	1034.8& 160.2& -0.107& 	1195.5&396.9& 	-0.131& 76.6\\
		25mJ*3times& 864.7& 188.9& 	-0.043& 	1037.3& 159.5& -0.098& 	1200.6&382.7& 	-0.115& 72.6\\
		25mJ*4times& 854.0& 193.9& 	-0.038& 	1037.1& 174.3& -0.089& 	1218.4&346.0& 	-0.089& 70.9\\
		25mJ*5times& 844.3& 296.3& 	-0.037& 	1039.9& 184.5& -0.065& 	1222.5&333.8& 	-0.059& 71.1\\
		\hline\noalign{\smallskip}
	\end{tabular}
	\ec
	\tablecomments{0.990\textwidth}{ The estimated Fo\# result is the average of the predicted values of the three diagnostic bands.}
\end{table}

The three diagnostic band centers of the original olivine spectrum deconvolved by MGM with Method P are 844.6 nm, 1025.9 nm, and 1197.9 nm, respectively. And the deconvolution results with Method E are 847.7 nm, 1027.9 nm, and 1201.6 nm, respectively. It is worth pointing out that the predicted Fo\# of these three band centers using the Method P regression equation in this paper are Fo79.4, Fo83.1, and Fo95.2, respectively. And the predicted Fo\# results with Method E are Fo78.5, Fo84.2, and Fo86.6, respectively. The average number of these two methods are Fo85.9 with Method P and Fo83.1 with Method E, respectively. There are  close to its chemical composition Fo89.8. However, applying the regression equation using in \cite{Sunshine1998Determining} with the deconvolution results of Method P, it will result in three inappropriate results, which are Fo83.9, Fo113.4, and Fo102.2, respectively . In addition, the average number is slightly larger than Fo99.8. Noticeably, when Fo\# is greater than 100, the predicted result is unacceptable and the estimated results differ from the chemical composition by more than 10 (mol\%). Therefore, it needs to be careful when using the regression equation based on \cite{Sunshine1998Determining}.

\begin{figure}
	\centering
	\includegraphics[height=8.5cm, width=12.0cm, angle=0]{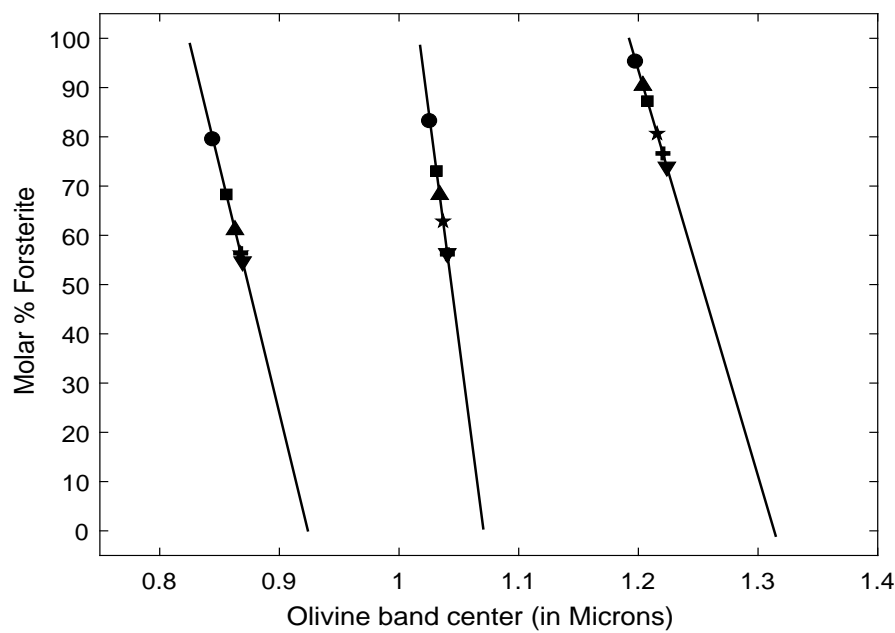}
	\caption{MGM fit to the original fresh and irradiated olivine spectra with Method P and predict the mole fraction of Fo\#. Black circle corresponds to original olivine sample, as the number of radiations increases, they correspond to black square, black upward-pointing triangle, black star, black downward-pointing triangle, and black cross, respectively.}
	\label{Fig13}
\end{figure}

Regarding to space weathering, after irradiations, these diagnostic spectral features turned narrower and shallower, and the three parameters (band center, band width, and band strength) of M1-1 band show the fastest change. The absorption features of M2 band are relatively stable, band centers gradually become larger and band strength gradually reduced, until the spectral absorption characteristics become very weak. The absorption intensity of the M1-2 band decays faster than the M2 band. Compared with the original olivine spectrum, all diagnostic band centers of the irradiated olivine spectra shift towards longer wavelength. Discuss in the most practical Method P, the band center offsets reach up to 24.7 nm, 14.5 nm, and 25.9 nm, for M1-1, M2, and M1-2 bands, respectively. This will cause the estimated Mg-numbers to be reduced from Fo85.9 to Fo61.6, corresponding to a change of more than 24 mol\% forsterite (or fayalite) in these olivine compositions. This result is consistent with the findings in \cite{fu2012effects}. Method E shows the same trend as Method P but with a different quantities. As a result, the predicted forsterite proportion on the space weathered asteroid surface based on the telescopic spectra may be less than its actual chemical composition.

\subsection{Estimating the composition of A-type asteroids (246) Asporina and (354) Eleonora}
\label{subsec:Estimating the A-type asteroids}

The regression equations based on the center of the olivine spectral features in this paper can be applied to estimate the surficial composition of olivine-dominated asteroids. However, the estimated chemical compositions can be changed by space weathering according to the above statement. Two A-type asteroids (246) Asporina and (354) Eleonora were chosen to study its possible mineralogical characterization. The VNIR spectra of (246) Asporina and (354) Eleonora (Fig. 14) were obtained from the website of Planetary Spectroscopy at MIT $\footnote{\url{http://smass.mit.edu/home.html}}$. The telescopic spectrum of (246) Asporina is a good sample for this study. It is inferred to be an olivine rich asteroid based on the dominated broad absorption feature near 1.0 nm in its VNIR spectrum. Whereas, the reddened spectrum has been interpreted to be affected by space weathering that produce metal particles on the surface (\citealt{cruikshank1984meteorite}). Spectrum of (354) Eleonora is in a similar situation (\citealt{Sunshine2007Olivine}).

\begin{figure}
	\centering
	\includegraphics[width=15.2cm, angle=0]{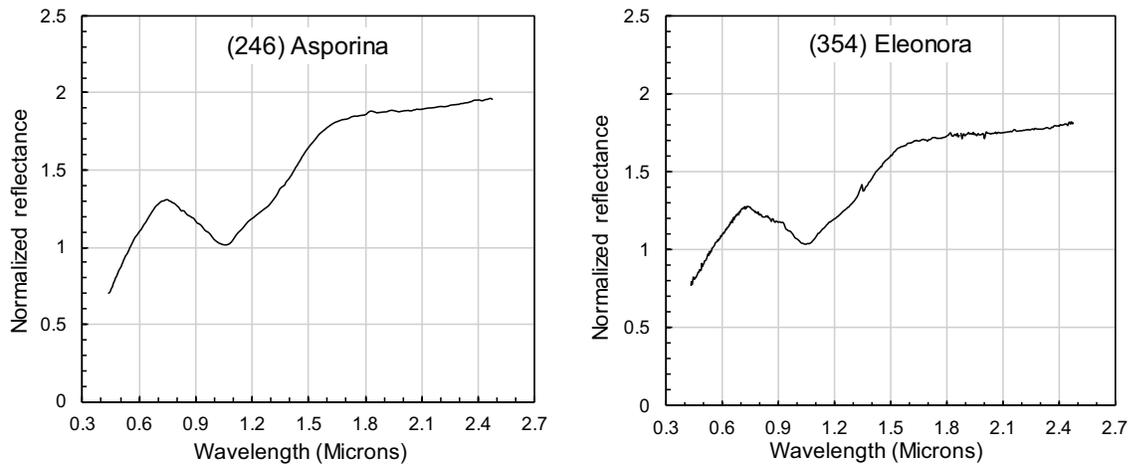}
	\caption{VNIR spectra of asteroid (246) Asporina and (354) Eleonora. The spectra is normalized to unity at 0.55 $\mu$m.}
	\label{Fig14}
\end{figure}

The forsterite mole percentages of asteroids (246) Asporina has been estimated to be between Fo60 and Fo90 by comparing the spectral curves of the olivine solid solution ranging from fayalite to forsterite (\citealt{cruikshank1984meteorite}). \cite{Sunshine1998Determining} also estimated the composition of olivine on asteroids (246) Asporina using the MGM with some restrictions and concluded that it was likely greater than Fo80. For (354) Eleonora, \cite{gaffey2015asteroid} reported that the Fo\# estimate is from $\sim$Fo61 to $\sim$Fo71, while \cite{Sunshine2007Olivine} modified the temperature and quantified its Fo\# to be $\sim$Fo90.

\begin{figure}
	\centering
		\begin{minipage}{7.2cm}
			\centering
			\includegraphics[width=7.4cm]{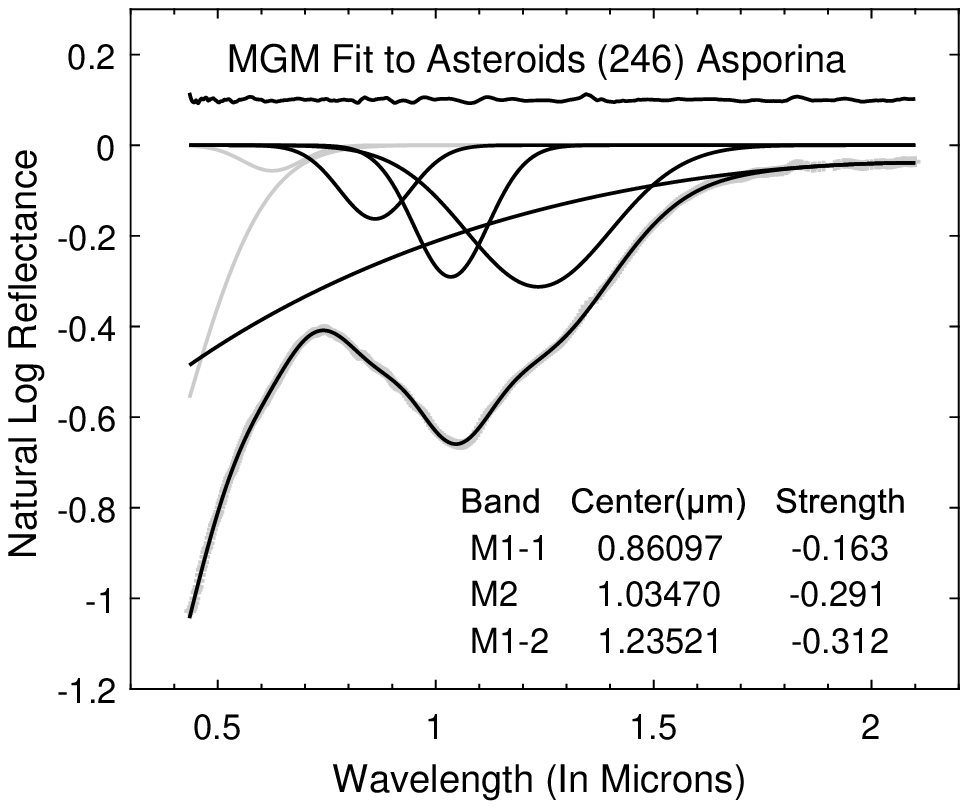}
		\end{minipage}
		\begin{minipage}{7.2cm}
			\centering
			\includegraphics[width=7.4cm]{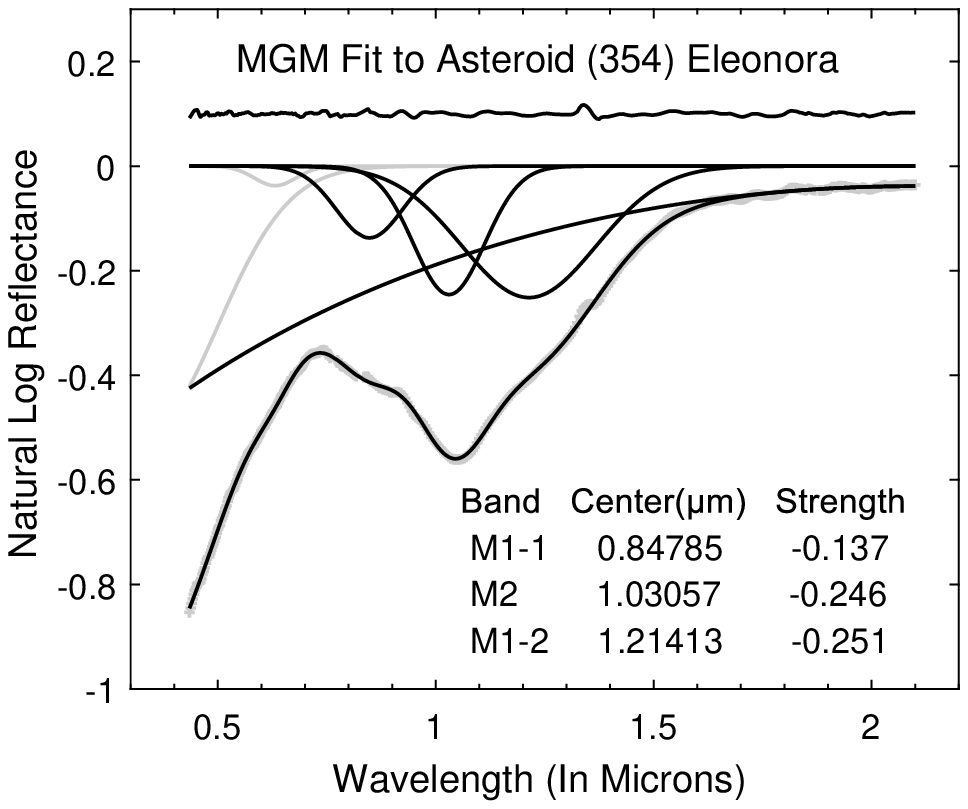}
		\end{minipage}
	\caption{MGM deconvolution results of A-type asteroids (246) Asporina and (354) Eleonora}
	\label{Fig15}
\end{figure}

This paper smoothed the asteroids spectra and normalized at the maximum value to acclimatize to the MGM. Then, MGM was applied to deconvolve the spectra with an appropriate second-order polynomial continuum, as shown in Fig. 15. The three diagnostic band centers of asteroid (246) Asporina spectrum are 860.97 nm, 1034.70 nm, and 1235.21 nm, respectively. The predicted Mg-numbers are Fo63.0, Fo66.8, and Fo64.5, respectively, and the average number is $\sim$Fo65. Whereas, the three diagnostic band centers of asteroid (354) Eleonora spectrum are 847.85 nm, 1030.57 nm, and 1214.13 nm, respectively, and the predicted Mg-numbers are Fo76.2, Fo74.4, and Fo81.9, respectively, the average number is $\sim$Fo78.

Based on the previous statement, space weathering may change the shape of spectral curves, and result in the underestimation of forsterite content. Based on equation (6), the $SSDF_{olivine}$ of asteroid (246) Asporina and (354) Eleonora are 0.3130 and 0.3105, respectively. Compared to the shape and $SSDF_{olivine}$ of the irradiated olivine spectra shown in Fig. 11 and Table 7, the spectral curves of the two asteroids are similar to the spectra of olivine irradiated with 25 mJ*1 time and 25 mJ*2 times. Moreover, the $SSDF_{olivine}$ values of the two asteroids are also in between the $SSDF_{olivine}$ values of these two irradiated olivine samples (0.4114 $\sim$ 0.2512). This means there may be about 10 mol\% forsterite underestimated. Therefore, considering the effects of space weathering on mineral chemical composition and the MGM deconvolution results, the Mg-numbers of asteroid (246) Asporina and (354) Eleonora are likely close to or greater than Fo75 and Fo88, respectively.

\section{Conclusions}
\label{sect:Conclusions}

MGM is a powerful tool to reveal the composition of mafic mineral assemblages. However, different continuum removal methods will affect the deconvolution results. This paper explored the continuum removal method for the MGM on olivine diagnostic spectral features. It used four different types of continuum removal methods to get the olivine diagnostic spectral features, which have resulted in four different regression equation systems. In comparison with different regression equations obtained in this article and \cite{Sunshine1998Determining}, it is significant to highlight the following:

1) Flat line continuum removal method is the original choice by \cite{Sunshine1998Determining}, it has many limitations in the deconvolution of olivine spectra. This method maybe overestimate the visible region if the flat continuum is far from the UV band wings, which makes it not suitable for revealing spectra with slopes;

2) Oblique line continuum removal method is simple and feasible, the tangent continua contact the two inflection regions of the diagnostic feature, this method is suitable for an individual absorption region, but not for the whole band;

3) A suitable second-order polynomial continuum removal method is also widely used, which is obviously superior to others. With this method, the overall shape of the continuum can well describe the general spectral trend;

4) Energy-wavelength polynomial continuum removal method is developed for deconvolving the reflectance spectra of lunar soils, this method does not perform well in laboratory spectra and the complex parameters are not friendly to the MGM program.

When using a suitable second-order polynomial continuum removal method, we get the optimized Mg-number regression equations as shown in Fig. 8 and Table 2. The validation and analysis of the deconvolution results of olivine-dominated Brachinite spectra indicate that the optimized Mg-number regression equations perform well for predicting the olivine composition in the laboratory. Furthermore, the effects of space weathering on the diagnostic spectral features of olivine are studied. The spectra of olivine irradiated with pulse laser for simulating varied degrees of space weathering were analyzed with the MGM. $SSDF_{olivine}$ function was developed to quantify the stability of the spectral diagnostic features of olivine and its degree of space weathering, and this function is independent of absolute reflectance. The deconvolution results of the irradiated olivine spectra and their $SSDF_{olivine}$ values show that the estimation for forsterite proportion on the surface of A-type asteroids based on the telescopic spectra may be less than its actual chemical composition. However, the underestimated amount is still difficult to be accurately determined. In conclusion, there is a need for more research on space weathering to improve the applicability and accuracy of the Modified Gaussian Model.

\normalem
\begin{acknowledgements}
	
This work was supported by the Foundation of the State Key Laboratory of Lunar and Planetary Sciences, Macau University of Science and Technology, Macau, China. The corresponding author X. Lu was also funded by The Science and Technology Development Fund, Macau SAR (File No. 0018/2018/A). Y. Yang was supported by Beijing Municipal Science and Technology Commission (File No. Z181100002918003). H. Zhang was supported by Natural Science Foundation of China (File No. U1631124, 11773023, and 11941001). This work was also supported by the grants from The Science and Technology Development Fund, Macau SAR (File No. 0007/2019/A). We are very grateful to Hu Xiaoyi, Ma Pei, and Jiang Te for their assistance in the experiments, whose comments and suggestions also improve the manuscript. Part of the spectral data utilized in this research were obtained and made available by the NASA RELAB Spectral Database at Brown University, Planetary Spectroscopy data at MIT, and USGS Spectral Library. The MGM program was developed and provided by Brown University. We acknowledge the support of Brown University, MIT, and USGS. And we thank the anonymous Reviewers and Editors for their assistance in evaluating this paper.

\end{acknowledgements}


\begin{thebibliography}{75}
\providecommand\natexlab[1]{#1}
\providecommand\JournalTitle[1]{#1}

\bibitem[Adams(1974)]{adams1974visible}
Adams, J.~B. 1974, J. Geophys. Res., 79, 4829

\bibitem[Antonenko \& Cloutis(2003)]{antonenko2003analysis}
Antonenko, I., \& Cloutis, E.~A. 2003, LPSC, 34

\bibitem[Basilevsky {et~al.}(2012)]{basilevsky2012geologic}
Basilevsky, A.~T., Shalygin, E.~V., Titov, D.~V., {et~al.} 2012, Icarus, 217,
  434

\bibitem[Binzel {et~al.}(2009)]{binzel2009spectral}
Binzel, R.~P., Rivkin, A.~S., Thomas, C.~A., {et~al.} 2009, Icarus, 200, 480

\bibitem[Bishop {et~al.}(1998)]{bishop1998spectroscopic}
Bishop, J.~L., Pieters, C.~M., Hiroi, T., \& Mustard, J.~F. 1998, MPSC, 33, 699

\bibitem[Brunetto {et~al.}(2006)]{brunetto2006space}
Brunetto, R., Romano, F., Blanco, A., {et~al.} 2006, Icarus, 180, 546

\bibitem[Burbine {et~al.}(2002)]{burbine2002spectra}
Burbine, T.~H., McCOY, T.~J., Nittler, L.~R., {et~al.} 2002, MPSC, 37, 1233

\bibitem[Burns(1970)]{burns1970crystal}
Burns, R.~G. 1970, American Mineralogist: Journal of Earth and Planetary
  Materials, 55, 1608

\bibitem[Burns(1974)]{burns1974polarized}
Burns, R.~G. 1974, American Mineralogist: Journal of Earth and Planetary
  Materials, 59, 625

\bibitem[Chapman(1996)]{chapman1996s}
Chapman, C.~R. 1996, MPSC, 31, 699

\bibitem[Chapman(2004)]{chapman2004space}
Chapman, C.~R. 2004, Annu. Rev. Earth Planet. Sci., 32, 539

\bibitem[Clark {et~al.}(2002)]{clark2002asteroid}
Clark, B.~E., Hapke, B., Pieters, C., \& Britt, D. 2002, Asteroids III, 585,
  90086

\bibitem[Clark \& Roush(1984)]{clark1984reflectance}
Clark, R.~N., \& Roush, T.~L. 1984, J. Geophys. Res: Solid Earth, 89, 6329

\bibitem[Clark {et~al.}(2003)]{clark2003imaging}
Clark, R.~N., Swayze, G.~A., Livo, K.~E., {et~al.} 2003, Journal of Geophysical
  Research: Planets, 108

\bibitem[Cl{\'e}net {et~al.}(2011)]{clenet2011new}
Cl{\'e}net, H., Pinet, P., Daydou, Y., {et~al.} 2011, Icarus, 213, 404

\bibitem[Cloutis \& Bell~III(2000)]{cloutis2000diaspores}
Cloutis, E.~A., \& Bell~III, J.~F. 2000, Journal of Geophysical Research:
  Planets, 105, 7053

\bibitem[Cruikshank \& Hartmann(1984)]{cruikshank1984meteorite}
Cruikshank, D.~P., \& Hartmann, W.~K. 1984, Science, 223, 281

\bibitem[De~Leon {et~al.}(2004)]{deLeon2004mineralogical}
De~Leon, J., Duffard, R., Licandro, J., \& Lazzaro, D. 2004, A\&A, 422, L59

\bibitem[Dyar {et~al.}(2009)]{dyar2009spectroscopic}
Dyar, M., Sklute, E., Menzies, O., {et~al.} 2009, American Mineralogist, 94,
  883

\bibitem[Fu {et~al.}(2012)]{fu2012effects}
Fu, X., Zou, Y., Zheng, Y., \& Ouyang, Z. 2012, Icarus, 219, 630

\bibitem[Gaffey(2010)]{gaffey2010space}
Gaffey, M.~J. 2010, Icarus, 209, 564

\bibitem[Gaffey {et~al.}(2015)]{gaffey2015asteroid}
Gaffey, M.~J., Reddy, V., Fieber-Beyer, S., \& Cloutis, E. 2015, Icarus, 250,
  623

\bibitem[Gallie {et~al.}(2008)]{gallie2008equivalence}
Gallie, E., Lyder, D., Rivard, B., \& Cloutis, E. 2008, International journal
  of Remote sensing, 29, 4089

\bibitem[Green {et~al.}(2011)]{green2011moon}
Green, R., Pieters, C., Mouroulis, P., {et~al.} 2011, J. Geophys. Res: Planets,
  116

\bibitem[Hapke(2001)]{hapke2001space}
Hapke, B. 2001, J. Geophys. Res: Planets, 106, 10039

\bibitem[Hiroi {et~al.}(2000)]{hiroi2000improved}
Hiroi, T., Pieters, C., \& Noble, S. 2000, LPSC

\bibitem[Hunt(1977)]{hunt1977spectral}
Hunt, G.~R. 1977, Geophysics, 42, 501

\bibitem[Hunt \& Ashley(1979)]{hunt1979spectra}
Hunt, G.~R., \& Ashley, R.~P. 1979, Economic Geology, 74, 1613

\bibitem[Isaacson {et~al.}(2011)]{isaacson2011remote}
Isaacson, P.~J., Pieters, C.~M., Besse, S., {et~al.} 2011, J. Geophys. Res:
  Planets, 116

\bibitem[Jin {et~al.}(2013)]{jin2013new}
Jin, S., Arivazhagan, S., \& Araki, H. 2013, Advances in Space Research, 52,
  285

\bibitem[Kohout {et~al.}(2014)]{kohout2014space}
Kohout, T., {\v{C}}uda, J., Filip, J., {et~al.} 2014, Icarus, 237, 75

\bibitem[Kokaly {et~al.}(2017)]{kokaly2017usgs}
Kokaly, R., Clark, R., Swayze, G., {et~al.} 2017, USGS Spectral Library Version
  7 Data: US Geological Survey data release

\bibitem[Li {et~al.}(2019)]{li2019chang}
Li, C., Liu, D., Liu, B., {et~al.} 2019, Nature, 569, 378

\bibitem[Lindsay {et~al.}(2015)]{lindsay2015composition}
Lindsay, S.~S., Marchis, F., Emery, J.~P., Enriquez, J.~E., \& Assafin, M.
  2015, Icarus, 247, 53

\bibitem[Lucey(2004)]{lucey2004mineral}
Lucey, P.~G. 2004, Geophys. Res. Lett., 31

\bibitem[Lucey {et~al.}(1998)]{lucey1998feo}
Lucey, P.~G., Taylor, G.~J., Hawke, B.~R., \& Spudis, P.~D. 1998, J. Geophys.
  Res: Planets, 103, 3701

\bibitem[Mittlefehldt {et~al.}(2003)]{mittlefehldt2003brachinites}
Mittlefehldt, D.~W., Bogard, D.~D., Berkley, J.~L., \& Garrison, D.~H. 2003,
  MPSC, 38, 1601

\bibitem[Miyazaki {et~al.}(2013)]{miyazaki2013olivine}
Miyazaki, T., Sueyoshi, K., \& Hiraga, T. 2013, Nature, 502, 321

\bibitem[Mustard {et~al.}(2005)]{mustard2005olivine}
Mustard, J.~F., Poulet, F., Gendrin, A., {et~al.} 2005, Science, 307, 1594

\bibitem[Nehru {et~al.}(1983)]{nehru1983brachina}
Nehru, C., Prinz, M., Delaney, J., {et~al.} 1983, J. Geophys. Res: Solid Earth,
  88, B237

\bibitem[Nehru {et~al.}(1992)]{nehru1992brachinites}
Nehru, C., Prinz, M., Weisberg, M., {et~al.} 1992, Meteoritics, 27

\bibitem[Nesvorn{\`y} {et~al.}(2009)]{nesvorny2009asteroidal}
Nesvorn{\`y}, D., Vokrouhlick{\`y}, D., Morbidelli, A., \& Bottke, W.~F. 2009,
  Icarus, 200, 698

\bibitem[Noble {et~al.}(2006)]{noble2006using}
Noble, S.~K., Pieters, C.~M., Hiroi, T., \& Taylor, L.~A. 2006, J. Geophys.
  Res: Planets, 111

\bibitem[Ody {et~al.}(2013)]{ody2013global}
Ody, A., Poulet, F., Bibring, J.-P., {et~al.} 2013, J. Geophys. Res: Planets,
  118, 234

\bibitem[Ohtake {et~al.}(2009)]{ohtake2009global}
Ohtake, M., Matsunaga, T., Haruyama, J., {et~al.} 2009, Nature, 461, 236

\bibitem[Pieters {et~al.}(1993)]{Pieters1993Optical}
Pieters, C.~M., Fischer, E.~M., Rode, O., \& Basu, A. 1993, J. Geophys. Res:
  Planets, 98, 20817

\bibitem[Pieters \& Mustard(1988)]{Pieters1988Exploration}
Pieters, C.~M., \& Mustard, J.~F. 1988, Remote Sensing of Environment, 24, 151

\bibitem[Pieters {et~al.}(2000)]{pieters2000space}
Pieters, C.~M., Taylor, L.~A., Noble, S.~K., Keller, L.~P., \& Wentworth, S.
  2000, MPSC Suppl., 35, A127

\bibitem[Pieters {et~al.}(2009)]{pieters2009moon}
Pieters, C.~M., Boardman, J., Buratti, B., {et~al.} 2009, Current Science, 500

\bibitem[Pinet {et~al.}(2009)]{Pinet2009Mafic}
Pinet, P.~C., Clenet, H., Heuripeau, F., {et~al.} 2009, LPSC

\bibitem[Rodger {et~al.}(2012)]{Rodger2012A}
Rodger, A., Laukamp, C., Haest, M., \& Cudahy, T. 2012, Remote Sensing of
  Environment, 118, 273

\bibitem[Roush {et~al.}(2015)]{Roush2015Laboratory}
Roush, T.~L., Bishop, J.~L., Brown, A.~J., Blake, D.~F., \& Bristow, T.~F.
  2015, Icarus, 258, 454

\bibitem[Salisbury \& Eastes(1985)]{Salisbury1985The}
Salisbury, J.~W., \& Eastes, J.~W. 1985, Icarus, 64, 586

\bibitem[Salisbury \& Wald(1992)]{SALISBURY1992The}
Salisbury, J.~W., \& Wald, A. 1992, Icarus, 96, 121

\bibitem[Savitzky \& Golay(1964)]{Savitzky1964Smoothing}
Savitzky, A., \& Golay, M. J.~E. 1964, Analytical Chemistry, 36, 1627

\bibitem[Shearer \& Papike(1999)]{Shearer1999Magmatic}
Shearer, C.~K., \& Papike, J.~J. 1999, American Mineralogist, 84, 1469

\bibitem[Singer(1981)]{Singer1981Near}
Singer, R.~B. 1981, J. Geophys. Res: Solid Earth, 86, 7967

\bibitem[Staid {et~al.}(2011)]{Staid2011The}
Staid, M.~I., Pieters, C.~M., Besse, S., {et~al.} 2011, J. Geophys. Res:
  Planets, 116, 239

\bibitem[Sugita {et~al.}(2011)]{Sugita2011A}
Sugita, S., Nagata, K., Tsuboi, N., Hiroi, T., \& Okada, M. 2011, LPSC

\bibitem[Sunshine {et~al.}(2007)]{Sunshine2007Olivine}
Sunshine, J.~M., Bus, S.~J., Corrigan, C.~M., Mccoy, T.~J., \& Burbine, T.~H.
  2007, MPSC, 42, 155

\bibitem[Sunshine {et~al.}(2004)]{Sunshine2004High}
Sunshine, J.~M., Bus, S.~J., Mccoy, T.~J., {et~al.} 2004, MPSC, 39, 1343

\bibitem[Sunshine \& Pieters(1993)]{Sunshine1993Estimating}
Sunshine, J.~M., \& Pieters, C.~M. 1993, J. Geophys. Res: Planets, 98, 9075

\bibitem[Sunshine \& Pieters(1998)]{Sunshine1998Determining}
Sunshine, J.~M., \& Pieters, C.~M. 1998, J. Geophys. Res: Planets, 103, 13675

\bibitem[Sunshine {et~al.}(1990)]{Sunshine1990Deconvolution}
Sunshine, J.~M., Pieters, C.~M., \& Pratt, S.~F. 1990, J. Geophys. Res: Solid
  Earth, 95, 6955

\bibitem[Tarantola \& Valette(1982)]{Tarantola1982Generalized}
Tarantola, A., \& Valette, B. 1982, Reviews of Geophysics, 20, 219

\bibitem[Taylor(1978)]{Taylor1978Geochemical}
Taylor, S.~R. 1978, in LPSC

\bibitem[Tonks \& Melosh(1992)]{Tonks1992Magma}
Tonks, W.~B., \& Melosh, H.~J. 1992, J. Geophys. Res: Planets, 98, 5319

\bibitem[Tsuboi {et~al.}(2010)]{Tsuboi2010A}
Tsuboi, N., Sugita, S., Hiroi, T., Nagata, K., \& Okada, M. 2010, in LPSC

\bibitem[Wang {et~al.}(2016)]{Wang2016Analysis}
Wang, H., Ma, Y., Zhao, H., \& Lu, X.-P. 2016, Acta Astronomica Sinica, 41, 419

\bibitem[Wilson \& Head(2017)]{wilson2017eruption}
Wilson, L., \& Head, J.~W. 2017, Journal of Volcanology and Geothermal
  Research, 335, 113

\bibitem[Wu {et~al.}(2010)]{Wu2010Global}
Wu, Y., Zhang, X., Yan, B., {et~al.} 2010, Science China(Physics,Mechanics \&
  Astronomy), 53, 2160

\bibitem[Yamamoto {et~al.}(2010)]{Yamamoto2010Possible}
Yamamoto, S., Nakamura, R., Matsunaga, T., {et~al.} 2010, Nature Geoscience, 3,
  533

\bibitem[Yang {et~al.}(2017)]{Yang2017Optical}
Yang, Y., Zhang, H., Wang, Z., {et~al.} 2017, A\&A, 597, A50

\bibitem[Zhang {et~al.}(2014)]{Zhang2014Effects}
Zhang, H., Yang, Y., Jin, W., Liu, C., \& Hsu, W. 2014, Optics express, 22,
  21280

\bibitem[Zhang {et~al.}(2016)]{Zhang2016Study}
Zhang, X., Ouyang, Z., Zhang, X., {et~al.} 2016, RAA, 16, 133

\end{thebibliography}

\end{document}